\documentclass[cernpreprint,texlive=2011,txfonts,UKenglish]{latex/atlasdoc} 
\pdfoutput=1

\usepackage{latex/atlaspackage}
\usepackage{latex/atlasbiblatex}
\usepackage{latex/tau3mu.paper}
\usepackage[percent]{overpic}
\usepackage{latex/slashbox}
\usepackage{pdflscape}

\graphicspath{{logos/}{fig/}}

\AtlasTitle{Probing lepton flavour violation via neutrinoless \boldmath$\tau\longrightarrow 3\mu$ decays with the ATLAS detector}

\PreprintIdNumber{CERN-PH-EP-2015-315}
\AtlasJournal{Eur. Phys. J. C.}

\author{The ATLAS collaboration}

\AtlasAbstract{
This article presents the sensitivity of the ATLAS experiment to the lepton-flavour-violating decays of \tmmm.
A method utilising the production of $\tau$ leptons via \Wtn decays is used.
This method is applied to the sample of 20.3~\ifb of $pp$ collision data at a centre-of-mass energy of 8~TeV collected by the ATLAS experiment at the LHC in 2012.
No event is observed passing the selection criteria, and the observed (expected) upper limit on the $\tau$ lepton branching fraction into three muons, \BRtmmm, is $3.76\times 10^{-7}$ ($3.94\times 10^{-7}$) at 90\% confidence level.
}

\begin{document}

\maketitle

\tableofcontents
\clearpage

\newcommand{\AtlasCoordFootnote}{%
ATLAS uses a right-handed coordinate system with its origin at the nominal interaction point (IP)
in the centre of the detector and the $z$-axis along the beam pipe.
The $x$-axis points from the IP to the centre of the LHC ring,
and the $y$-axis points upwards.
Cylindrical coordinates $(r,\phi)$ are used in the transverse plane, 
$\phi$ being the azimuthal angle around the $z$-axis.
The pseudorapidity is defined in terms of the polar angle $\theta$ as $\eta = -\ln \tan(\theta/2)$.
Angular distance is measured in units of $\Delta R \equiv \sqrt{(\Delta\eta)^{2} + (\Delta\phi)^{2}}$.}

\section{Introduction}
\label{sec:intro}
The observation of a lepton-flavour-violating (LFV) process involving charged leptons would be a major breakthrough in understanding the matter content of the universe and would support the hypothesis of leptogenesis~\cite{Davidson:2008bu}.
In particular, LFV processes involving both a $\tau$ lepton and a muon are seen as most promising for such an observation, given the current measurements of neutrino oscillations~\cite{PDG}.
In the Standard Model (SM), such processes have a vanishingly small branching fraction, e.g. \BRtmmm $<10^{-14}$~\cite{Pham:1998fq}, while a number of models beyond the SM predict it to be of the order of $10^{-10}$--$10^{-8}$~\cite{FlavourPhysics:2008,Abada:2014kba,Arganda:2005ji}.
The current limits on branching fractions of neutrinoless $\tau$ lepton decays are of the order of few times $10^{-8}$~\cite{Babar:2010ez,BelleLimit:2007,LHCb:2014,HFAG:2014}, for $Z$ boson LFV decays they are about $10^{-5}$~\cite{PDG,DELPHIZtaumu,OPALZtaue}, and for the LFV decay of a Higgs boson to a $\tau$ lepton and a muon they are about 1\%~\cite{Khachatryan:2015kon,Aad:2015gha}.
The main experimental obstacles to improve the sensitivity with $\tau$ leptons are the small number of produced $\tau$ leptons world-wide.

In this article, a search for neutrinoless $\tau$ lepton decays to three muons is performed with 20.3~\ifb of $pp$ collision data collected with ATLAS detector in 2012 at 8~TeV centre-of-mass energy.
The search is focused on a particular source of $\tau$ leptons, namely \Wtn decays with subsequent \tmmm decay.
In such events, $\tau$ leptons are produced with a transverse momentum (\pT) mostly in the range of $\sim 25-50$~GeV.
Due to the relativistic boost of the $\tau$ lepton, the muons from the $\tau$ LFV decay are produced in close geometrical proximity to each other but isolated from other energetic particles in the event.
The tau-neutrino from the $W$ boson decay appears as missing transverse momentum (\etmis) in the detector and together with the transverse momentum of the three muons ($p_{\rm T}^{3\mu}$) gives a transverse mass, $m_{\rm T}=\sqrt{2p_{\rm T}^{3\mu}\etmis(1-\cos\Delta\phi)}$, compatible with the $W$ boson decay, where $\Delta\phi$ is the angle between the directions of the $p_{\rm T}^{3\mu}$ and the \etmis.
The unique signature in the detector is three muons with invariant mass equal to the mass of the $\tau$ lepton and with a significant missing transverse momentum that is on average back-to-back with the three muons in the transverse plane.
Since no energetic jet is expected in the majority of $W$ boson production events, very small hadronic activity is predicted beyond that from the soft underlying event or multiple simultaneous $pp$ collisions (pile-up).
A large fraction of such $\tau$ leptons decay sufficiently far from the $W$ production vertex to give a fully reconstructable additional vertex.
This allows the selection of three muons originating from a vertex which is displaced from the primary interaction vertex.
The background events usually contain one or two muons originating from the decay of hadrons, including decays in flight, while the remaining tracks are hadrons mis-measured as muons, originating from e.g. a pile-up jet or a pion punching through the calorimeter.
The dominant background is due to muons originating from decays of $b$- or $c$-hadrons (heavy flavour, HF).
Although such decays are typically accompanied by jets of particles produced in the direction opposite to the HF jets, in a fraction of the events the associated jet is lost or mis-measured, mimicking the signal \etmis.
A small light-flavour multi-jet contribution is also present while the contribution from leptonic decays of vector bosons is negligible.

The analysis strategy is as follows.
Events with three muons associated with a common vertex are selected.
A loose event selection is applied to collect a high-quality sample of candidate events satisfying $|\mthreebody-m_\tau |\lesssim 1$~GeV.
The characteristics of the loose sample of events are then analysed with a {\it boosted decision tree} (BDT).
The BDT input variables are chosen so that the BDT output and the three-muon mass are uncorrelated in the mass range used in the analysis.
A tight selection, following an initial cut on the BDT output, is applied to separate the signal from the background.
After the optimal cut on the BDT output is found, a search is performed for an excess of events at the $\tau$ lepton mass above the expected background level.

The branching fraction is calculated as
\begin{equation}
	\BRtmmm = \frac{\Nsig}{\left(\acceff\right)\NWtn},
\label{eq:BR}
\end{equation}
where \Nsig is the number of observed events above the expected background level in a narrow region around the $\tau$ lepton mass, \acceff is the detector acceptance times efficiency for the signal, and \NWtn is the number of $\tau$ leptons produced via the \Wtn channel (additional contributions to the $\tau$ lepton yield are estimated to be less than 3\%).

\section{The ATLAS detector}
\label{app:atlasdet}

The ATLAS experiment~\cite{PERF-2007-01} at the LHC is a multi-purpose particle detector
with a forward-backward symmetric cylindrical geometry and a near $4\pi$ coverage in solid angle.\footnote{\AtlasCoordFootnote}
It consists of an inner tracking detector surrounded by a thin superconducting solenoid providing a \SI{2}{\tesla} axial magnetic field, electromagnetic and hadronic calorimeters, and a muon spectrometer.
The inner tracking detector (ID) covers the pseudorapidity range $|\eta | < 2.5$.
It consists of silicon pixel, silicon microstrip, and transition radiation tracking detectors.
Lead/liquid-argon (LAr) sampling calorimeters provide electromagnetic (EM) energy measurements with high granularity.
A hadronic (iron/scintillator-tile) calorimeter covers the central pseudorapidity range ($|\eta | < 1.7$).
The endcap and forward regions are instrumented with LAr calorimeters for EM and hadronic energy measurements up to $|\eta | = 4.9$.

The muon spectrometer (MS) comprises separate trigger and high-precision tracking chambers measuring the deflection of muons in a magnetic field generated by superconducting air-core toroids.
The magnets' bending power is in the range from \num{2.0} to \SI{7.5}{\tesla\metre}.
The muon tracking chambers cover the region $|\eta | < 2.7$ with three layers of monitored drift tubes, complemented by cathode-strip chambers in the forward region, where the background is highest.
The muon trigger system covers the range $|\eta | < 2.4$ with resistive-plate chambers in the barrel, and thin-gap chambers in the endcap regions.

A three-level trigger system is used to select events.
The first-level trigger is implemented in hardware and uses a subset of the detector information to reduce the accepted rate to at most \SI{75}{\kilo\hertz}.
This is followed by two software-based trigger levels that together reduce the accepted event rate to \SI{400}{\hertz} on average.

During the data-taking period, there were no dedicated triggers implemented for this analysis.
A combination of seven muon triggers is used, where all triggers are constructed from at least two trigger objects.
A detailed discussion of the trigger is given in Section~\ref{sec:trig_and_reco}.

\section{Simulation and data samples}
\label{sec:data_mc}

The results presented here are based on proton--proton collision data at a centre-of-mass energy of $\sqrt{s} = 8$~TeV, collected by the ATLAS detector at the LHC during 2012.
Data samples corresponding to an integrated luminosity of 20.3~\ifb are used.
Selected data events are required to have all relevant components of the ATLAS detector in good working condition.

The Monte Carlo (MC) simulated $W\to\tau\nu\to (3\mu)\nu$ signal sample is produced by the \pythia~\cite{Sjostrand:2014zea} event generator (version 8.175) using the AU2~\cite{AUTUNE} set of tuned parameters and the MSTW2008LO parton distribution function (PDF) set~\cite{AUMSTW2008LO}.
This signal sample is modelled using \Wtn production where the $\tau$ lepton is forced to decay isotropically into three muons as in previous searches for this mode~\cite{Babar:2010ez,BelleLimit:2007,LHCb:2014,HFAG:2014}.
The detector response is modelled using GEANT4~\cite{Agostinelli:2002hh,Aad:2010ah}.
The number of $\tau$ leptons produced in the 2012 dataset via the \Wtn channel appearing in Eq.~(\ref{eq:BR}), is estimated by scaling the ATLAS measurement of the $W\to\ell\nu$ cross-section at $\sqrt{s}=7$~TeV~\cite{Wlnucross} to 8~TeV using the ratio of the 8~TeV to 7~TeV NNLO cross-section calculations ($\sigma_{\rm theory}^{\rm 8~TeV} = 12.18\pm 0.61$~nb and $\sigma_{\rm theory}^{\rm 7~TeV} = 10.46\pm 0.52$~nb) and multiplying by the 8~TeV integrated luminosity.
The result is $\NWtn=(2.41\pm 0.08) \times 10^8$, taking into account the uncertainty reported in Ref.~\cite{Wlnucross} and the uncertainty in the 7 and 8~TeV luminosities.
For the selection applied in the analysis, the contamination from other sources of $\tau$ leptons, such as $Z\to\tau\tau$ or HF processes, is less than 3\% and is therefore neglected.
The background is estimated using data as discussed in Section~\ref{sec:background_estimation}.

\section{Trigger and reconstruction}
\label{sec:trig_and_reco}
To maximise the signal acceptance times efficiency, events are required to pass at least one of seven triggers.
These are six multi-muon triggers and one dimuon plus \etmis trigger.
The software-based trigger thresholds used for the muons range from 4 to 18~GeV in transverse momentum while the \etmis threshold is 30~GeV.
The trigger efficiency for simulated signal events within the muon-trigger acceptance (three generator-level muons with $\pT>2.5$~GeV and $|\eta |<2.4$) is $\sim 31\%$ for the combination of all triggers used in the analysis.
To evaluate the trigger performance in the region where the muons have a small angular separation, as is typical for the signal, a tag-and-probe study is performed using data events containing high-momentum \Jpsimm candidates.
For this study, the data are collected using a single-muon baseline trigger with a \pT threshold of 18~GeV.
Single-muon efficiencies are measured separately for the different thresholds which define the six multi-muon triggers.
Each multi-muon trigger efficiency is calculated as the product of the single-muon efficiencies.
Correction factors are applied to account for the limited performance of the trigger system in identifying a pair of muons as two muon-trigger objects.
At small angular separations ($\dR \lesssim 0.2$), where most of the signal is expected and where these limitations are most pronounced, these corrections must be taken into account.
These factors are measured from the efficiency to identify two independent muon-trigger objects for different \dR values between the tag- and the probe-muon.
The total efficiency of the seven triggers is calculated considering correlations between any of the triggers.
The trigger efficiency, measured from the data, is compared to the one measured in simulated \Jpsi events for the seven different triggers separately and jointly.
Agreement between data and MC simulation was found to be within 11\% for all relevant values of \dR and \pT, where the largest difference comes from events where the \dR separation is smallest.
The systematic uncertainty on \acceff due to the trigger is therefore taken to be 11\%.

The approach for measuring the muon reconstruction efficiency is similar to that used to measure the trigger efficiency.
While the trigger efficiency is measured with respect to muon reconstruction as the baseline, the reconstruction efficiency is measured with respect to ID tracking, which in turn is close to $100\%$ efficient~\cite{Aad:2014rra}.
Small deviations from the assumed value for ID tracking efficiency have a negligible impact on this measurement.
The tag-and-probe procedure is performed using muons as tags and ID tracks as probes. 
The baseline sample for the reconstruction efficiency measurement includes a large number of non-muon tracks, which must be subtracted.
This is done in bins of probe-track \pT ($p_{\rm T}^{\rm trk}$) and bins of the angular separation between the tag-muon and the probe-track, $\Delta R_{\mu + {\rm trk}}$.
To describe the \Jpsi peak and the background, a small range in tag-muon plus probe-track invariant mass, $m_{\mu+{\rm trk}}\in [2600,3500]$~MeV, is fit to a double Gaussian function plus an exponential function and a second-order polynomial.
In each $p_{\rm T}^{\rm trk}$ or $\Delta R_{\mu + {\rm trk}}$ bin, the ratio of the \Jpsi peak component integral to the full shape (\Jpsi plus background) integral is used as a weight to correct the $p_{\rm T}^{\rm trk}$ or $\Delta R_{\mu + {\rm trk}}$ shape itself. 
This is done separately for the probe-track distributions (denominators) and the muon-matched probe-track distributions (numerators).
The ratio of the above two weighted distributions is defined as the reconstruction efficiency per $p_{\rm T}^{\rm trk}$ or $\Delta R_{\mu + {\rm trk}}$ bin.
The efficiency measured with this approach in data is compared with the one from simulation and the difference at small $\Delta R_{\mu + {\rm trk}}$ results in an uncertainty of 13.1\% per event.

\section{Analysis procedure}
\label{sec:analysis_procedure}

The analysis procedure is divided into four steps. 
First, events containing three high-quality muon objects with a combined invariant mass of less than 2.5~GeV are selected.
These muons are required to originate from a common vertex.
Second, a \loose selection is applied to this sample to obtain a background sample that can be used to train the BDT, which is constructed using the TMVA toolkit~\cite{TMVA}.
The \loose selection cuts (using a number of vertex quantities as well as kinematic quantities) are chosen to obtain a large background sample for training, while rejecting background that is kinematically inconsistent with the signal.
Before training the BDT, the data events are divided into three regions based on the three-muon mass value.
These are the blinded region (which includes the signal region), a sideband region and a BDT training region as defined in Table~\ref{table:ranges}.
Third, a \tight selection (tightening the \loose selection with a few additional cuts) is applied while simultaneously placing an initial loose cut on the BDT score, denoted by \xzero.
The \xzero cut removes background-like events having a very low BDT score, while the \tight selection further reduces the background in the blinded and sideband regions.
Fourth, the background rejection as a function of the BDT cut is studied using data events in the sideband region passing the \tightxzero selection.
This allows to optimise the final cut on the BDT score, denoted by \xone.
The statistical analysis is performed for the \tightxone selection.

\begin{table}[!ht]
\begin{center}
\begin{tabular}{ll}
\hline
\hline
Region            & Range in \mthreebody [MeV]    \\
\hline                           
Signal region     & \SignalRegionBrackets  \\
Blinded region    & \BlindedRegionBrackets \\
Sideband region   & \SidebandLeftBrackets and \SidebandRightBrackets \\
Training region   & \TrainingLeftBrackets and \TrainingRightBrackets  \\
\hline 
\hline
\end{tabular}
\end{center}
\caption{The different three-muon mass ranges used in the analysis.}
\label{table:ranges}
\end{table}

The signal region (SR) is defined as an interval around the $\tau$ lepton mass with a half-width corresponding to twice the resolution of the three-muon mass, $\sigma_s = 32$ MeV, as obtained from the signal MC sample.
The analysis was blinded in a slightly wider region to allow variation of the signal region definition.
The signal MC sample is divided into two independent samples.
One signal sample is used for the BDT training while the second signal sample is used for estimating the \acceff.
The background in the signal region is estimated from a fit to the three-muon mass distribution in the sidebands (SB) using the \tightxzero selection.
This estimate is then scaled down to the final BDT score cut, $x_1$, using a fit to the BDT shape as explained below.

\subsection{Object selection}
\label{sec:object_selection}
Muons are selected to have a transverse momentum greater than 2.5~GeV and are required to pass stringent requirements on the track quality and the associated hits in both the ID and the MS.
Only combined ID+MS measurements of track parameters are used.
Several matching criteria~\cite{Aad:2014rra} are imposed to reject non-muon tracks (e.g. tracks from hadron decays in flight).
The performance of muon identification is validated in two dedicated dimuon control regions.
One region is populated with muons from $\Jpsi\to\mu\mu$ decay (in $2850 < \mtwobody < 3350$~MeV), while the second region has an enhanced fraction of non-muon tracks (in events with $\mtwobody < 750$~MeV).

Events with at least three selected muons are considered.
All possible three-muon combinations are used as inputs to a vertex fit.
The primary vertex (PV) is also refitted after removing the three tracks.
Due to the $\tau$ lepton lifetime, the three-muon vertex is often separated from the PV.
The characteristics of the separation between the three-muon vertex and the PV are therefore used to distinguish signal from background.
Particularly, the two projections of the three-muon vertex displacement with respect to the PV in the transverse plane are used; $\Lxy{=}L_{\rm T}\cos\theta_{xy}$ and $\azeroxy{=}L_{\rm T}\sin\theta_{xy}$ where $L_{\rm T}$ is the transverse component of the vector connecting the PV and the three-muon vertex and $\cos\theta_{xy}{=}\frac{\vec{L}_{\rm T}\cdot\vec{p}_{\rm T}^{3\mu}}{L_{\rm T} \pTthreebody}$.
The three-muon vertex fit probability, \pvalue, is also used (as calculated from the vertex fit $\chi^2$ and degrees of freedom).
After fitting all possible vertices, exactly one three-muon candidate is allowed per event, satisfying $\mthreebody<2500$~MeV and $|\qthreebody |=1$ where \qthreebody is the sum of the charges of the three-muon tracks.

Jets are used to separate the signal from the multi-jet backgrounds (predominantly HF), where more hadronic activity is expected.
The jets are reconstructed from topological clusters formed in the calorimeter using the anti-$k_t$ algorithm~\cite{Cacciari:2008gp} with a radius parameter $R=0.4$.
The jets are calibrated to the hadronic energy scale using energy- and $\eta$-dependent correction factors derived from simulation and with residual corrections from in situ measurements.
A detailed description of the jet energy scale measurement and its systematic uncertainties can be found in Ref.~\cite{Aad:2014bia}.
Jets found within a cone of $\Delta R = 0.2$ around a selected three-muon candidate are removed.
Jets are required to have $\pT>30$~GeV and $|\eta |<2.8$; only the leading jet satisfying these criteria is considered.
There is no veto of events with more than one jet satisfying these criteria.
The leading jet and the three-muon momenta are summed vectorially to define $\vH = \vec{p}_{\rm jet} + \vec{p}_{3\mu}$ with \HT being the magnitude of its transverse component.
For events where there are no jets satisfying these criteria (the majority of events for the signal), \vH is simply $\vec{p}_{3\mu}$.

The \etmis is calculated as the negative vector sum of the transverse momenta of all high-\pT objects reconstructed in the event, as well as a term for other activity in the calorimeter~\cite{Aad:2012re}.
Clusters associated with electrons, hadronic $\tau$ lepton decays and jets are calibrated separately, with other clusters calibrated at the EM energy scale.
This \etmis is denoted hereafter by \calomet.
In addition, a track-based missing transverse momentum (\trackmet) is calculated as the negative vector sum of the transverse momenta of tracks with $|\eta |<2.5$, $\pT>500$~MeV and associated with the primary vertex.
Both the calorimeter-based and track-based measurements of the \etmis are used.

Several kinematic variables are defined from the reconstructed objects listed above.
Two transverse masses are defined using the three-muon transverse momentum (\pTthreebody) as $m_{\rm T} = \sqrt{2\pTthreebody\etmis(1-\cos\Delta\phi_{3\mu})}$, where $\Delta\phi_{3\mu}$ is the angle between the \etmis and \pTthreebody directions in the transverse plane.
In these definitions, \etmis can be either \calomet or \trackmet to obtain \mTcalo or \mTtrack respectively.
The $\Delta\phi_{3\mu}$ terms are \dphithreebodymetcalo or \dphithreebodymettrack respectively.
Similarly, the $\Delta\phi_{\HT}$ variable is the angle between the \etmis and \HT directions in the transverse plane.
This adds two additional angles, \dphiHmetcalo and \dphiHmettrack for \calomet and \trackmet respectively.
These $\Delta\phi_{\HT}$ variables provide good separation when a hard jet is found and thus \HT deviates from \pTthreebody in magnitude and direction.

\subsection{Loose event selection}
\label{sec:loose_event_selection}
After the three-muon candidates are formed from the selected muons, a \loose event selection is performed, maintaining a signal efficiency of about 80\% while rejecting about 95\% of the background.
This \loose selection includes cuts on the displacement of the vertex from the PV, requirements on the three-muon kinematics and on the presence of other tracks (track isolation), and requirements on quantities involving \calomet and \trackmet.
The \loose selection comprises the following requirements:
\begin{itemize}
	\item The \Lxy significance, $\SLxy{=}\Lxy/\dLxy$, must satisfy $-10{<}\SLxy{<}50$, where \dLxy is the uncertainty in the \Lxy. 
	\item The \azeroxy significance, $\Sazeroxy{=}\azeroxy/\dazeroxy$, must satisfy $\Sazeroxy{<}25$, where \dazeroxy is the uncertainty in \azeroxy.
	\item The three-muon track-fit probability product, $\ptrks = p_1\times p_2\times p_3$ (where $p_i$ is the track fit \pvalue of track $i$), must satisfy $\ptrks>10^{-9}$.
	\item The three-muon transverse momentum must satisfy $\pTthreebody>10$~GeV.
	\item The missing transverse energies, \calomet and \trackmet, must both satisfy $10<\etmis< 250$~GeV.
	\item The transverse masses, \mTcalo and \mTtrack, must both satisfy $m_{\rm T}>20$~GeV.
	\item The three-muon track isolation is obtained from the sum of the \pT of all tracks with $p_{\rm T}^{\rm trk}>500$~MeV in a cone of $\dRmax+0.20$ (and $\dRmax+0.30$) around the three-muon momentum while excluding its constituent tracks; it must satisfy $\isoTwenty<0.3$ (and $\isoThirty<1$). The largest separation, $\dRmax$, between any pair of the three-muon tracks is on average 0.07 for the signal.
\end{itemize}
The loose cuts on the significances, \SLxy and \Sazeroxy, are applied to allow the three-muon vertex to be separated from the PV, while still being compatible with the $\tau$ lepton lifetime.
The requirement on \ptrks imposes a goodness-of-fit criterion on the three-muon candidate.
This value is based on an examination of signal-like events found in the sideband region in the data.
As this is not the only quality requirement imposed on the individual muon objects, it is kept loose in this part of the selection.
The efficiency for this cut to select signal events is $\sim 98\%$, while it is rejecting $\sim 13\%$ of the background events.
The kinematic and the isolation variables are very effective in separating the $W$ boson properties of the signal from the HF and the light-flavour multi-jet background, which tend to be non-isolated and with low values of \pT, \etmis and $m_{\rm T}$.
The associated cuts remain very loose in this part of the selection to ensure that the sample sizes are large enough for the BDT training.

\subsection{Multivariate analysis}
\label{sec:mva_setup}
The events passing the \loose selection described above are used as input to the BDT training.
There are 6649 events passing the loose selection in the signal MC sample (out of $10^5$), where 6000 of these events are used for the BDT training and the rest are used for testing the BDT output.
Similarly, the number of data events passing the \loose selection in the training region is 4672, where 4000 of these events are used for the BDT training.
The BDT input variables include kinematic distributions, track and vertex quality discriminants, vertex geometry parameters, and isolation.
The following variables (sorted by their importance ranking) are used as inputs to the BDT:
\begin{enumerate}
	\item The calorimeter-based transverse mass, \mTcalo.
	\item The track-based missing transverse momentum, \trackmet.
	\item The isolation variable, \isoTwenty.
	\item The transverse component of the vector sum of the three-muon and leading jet momenta, \HT.
	\item The track-based transverse mass, \mTtrack.
	\item The difference between the \calomet and \trackmet directions, \dphimets.
	\item The calorimeter-based missing transverse momentum, \caloetmis.
	\item The track-based missing transverse momentum balance $\pTthreebody/\tracketmis-1$.
	\item The difference between the three-muon and \caloetmis directions, \dphithreebodymetcalo.
	\item Three-muon vertex fit probability, \pvalue.
	\item The three-muon vertex fit \azeroxy significance, \Sazeroxy.
	\item The track fit probability product, \ptrks.
	\item The three-muon transverse momentum, \pTthreebody.
	\item The number of tracks associated with the PV (after refitting the PV while excluding the three-muon tracks), \pvntrk.
	\item The three-muon vertex fit \Lxy significance, \SLxy.
	\item The calorimeter-based missing transverse momentum balance, $\pTthreebody/\caloetmis-1$.
\end{enumerate}
This configuration was found to give the optimal balance between background rejection and signal efficiency.

The \HT variable is introduced to avoid vetoing events with at least one jet fulfilling the requirements listed in section~\ref{sec:object_selection}.
Although the majority of signal events do not have jets, it is found that keeping such events increases the \acceff by $\sim 15\%$ and also  ultimately leads to better rejection power, owing to the significantly larger training and sideband samples.
The variables \dphimets, \dphithreebodymetcalo, $\pTthreebody/\tracketmis-1$, $\pTthreebody/\caloetmis-1$ and \pvntrk are complementary to the \etmis-related variables used in the \loose selection as well as here.
These variables are also very effective in distinguishing the \Wtn production of the signal from the HF and light-flavour multi-jet background.
The vertex \pvalue is a variable complementary to the \SLxy and \Sazeroxy variables used in the \loose selection as well as here.
The HF and light-flavour multi-jet backgrounds have mostly random combinations of selected muon objects which do not originate from the same vertex.
This variable peaks at very low values for the background while for the signal it is distributed uniformly and thus provides excellent separation.

After training the BDT with data events from the training region and signal MC events from the first signal MC sample, the BDT response is calculated for the data events in the sidebands and for events in the second signal sample.
The BDT score, $x$, ranges between -1 and +1.
Events with a very low BDT score, within $-1\leq x\leq -0.9$ are removed from further consideration, defining $x_0\equiv -0.9$.

In order to assess potential modelling problems in the signal MC sample, the BDT input distributions and the BDT response are validated against single-muon data.
These data contain mainly \Wmn events with a small fraction ($<10\%$) of background.
The single-muon selection is formulated to be as close as possible to the main analysis selection where the differences are mostly driven by the different triggers used (one single-muon trigger with no isolation requirement and with a threshold of 24~GeV is used in the validation) and the exclusion of variables which do not have equivalents in the \Wmn case, e.g. the three-muon vertex variables.
The training samples used for this validation study, for both data and signal, are the same samples as used in the main analysis, constructed with the same \loose selection as described in the previous section.
All input variables are used for the training, excluding the \pvalue, \SLxy, \Sazeroxy and \ptrks, which cannot be calculated in a single-muon (\Wmn) selection.
The resulting BDT setup is hereafter referred to as ``partial BDT''.
After training the partial BDT, the response is tested on the second signal sample and on the single-muon data, using the single-muon selection and where the three muon objects in the signal sample are treated as one object (muon).
The \pvntrk distribution of the signal sample is also modified by subtracting two tracks to reflect the difference with respect to a single-muon selection.
The responses in data and simulation are compared and are found to agree within 10\% throughout most of the phase-space for all variables.
The ratio of the partial BDT responses for the single-muon data and signal MC events is used as an event weight while applying the full selection and calculating the weighted \acceff as described in the next sections.
The difference between the weighted and unweighted \acceff is found to be 4\% and is taken as a modelling uncertainty.

Any variable which may bias the BDT response by only selecting events very close to the $\tau$ lepton mass is not included in the BDT input list.
The distribution of the three-muon mass has been examined in several bins of $x$ above $x_0$ using both the \loose and the \tight samples, where no hint of potential peaking background around the $\tau$ lepton mass has been found.
In addition, the shape of the three-muon mass distribution has been found to be insensitive to the BDT cut, as expected given the small correlation coefficient between $x$ and \mthreebody, which is found to be about -0.05.

\subsection{Tight event selection}
\label{sec:tight_event_selection}
Additional \tight cuts are applied after the BDT training and the application of the \xzero cut on the BDT score.
The following requirements are tightened or added:
\begin{itemize}
	\item A number of the \loose requirements are tightened, namely $\ptrks>8\times10^{-9}$, $\mTcalo>45$~GeV, $\mTtrack>45$~GeV and $1{<}\SLxy{<}50$.
	\item Three-muon vertex fit probability must have $\pvalue>0.2$.
	\item The angle between the \HT and \calomet (\trackmet) directions is required to be $\dphiHmetcalo>2$ ($\dphiHmettrack>2$).
	\item The same-charge two-muon mass, \mSS, and opposite-charge two-muon mass, \mOSa or \mOSb, satisfy
	$\mSS>300$~MeV, $\mOSa>300$~MeV and $\mOSb>300$~MeV, where \mOSa (\mOSb) is the mass of the two opposite-charge muon pairs with the highest (second highest) summed scalar \pT among the three muons.
	\item The event is rejected if $|\mOS-\momega|<50$~MeV or $|\mOS-\mphi|<50$~MeV if either of the \pTthreebody, the \calomet or the \trackmet is lower than 35~GeV.
	\item The event is rejected if $|\mOS-\mphi|<50$~MeV if $|\mthreebody-\mDs|<100$~MeV.
\end{itemize}
In the above notation, \mOS is \mOSa or \mOSb and \momega, \mphi and \mDs are the masses of the $\omega$, $\phi$ and \Ds mesons respectively, taken from Ref.~\cite{PDG}.

The requirement on the three-muon vertex fit probability is applied in order to ensure a high-quality fit.
The cuts on \dphiHmetcalo (\dphiHmettrack) are applied in order to further suppress the HF and multi-jet background where the three-muon candidate is typically produced within or near a jet.

The first two-muon mass requirement is applied to suppress candidates originating from one prompt muon object and two muons from a converted photon.
The second requirement on the two-muon masses is applied to prevent the low-mass mesons, \rhoomega and $\phi$, from entering into the region close to the $\tau$ lepton mass when combined with an additional track.
In the selected three-muon event sample, these resonances appear as two clear peaks in the mass distribution of oppositely charged muon pairs in data.
Since the resonances lie in the middle of the signal distribution, the low-\pT and \etmis requirement ensures that these can still be distinguished from the signal, and thus it removes the resonances while still maintaining a high enough signal efficiency.
Finally, the last requirement is applied to remove a potential $\Ds\to\pi+\phi(\mu\mu)$ contamination from the high-mass sideband.
The cuts listed above comprise the \tight selection where the \tightxzero selection is used to estimate the background for any cut on $x$ above $x_0$.

Figure~\ref{fig:mass_and_score} shows the three-muon mass distribution and the BDT response distribution.
Figures~\ref{fig:BDTinputs1} and~\ref{fig:BDTinputs2} show the distributions of the BDT inputs sorted by the separation rank as reported by TMVA during the BDT training.
Figure~\ref{fig:other} shows the distributions of the complementary variables which are used in the \loose or \tight selection but not in the BDT.

\begin{figure}[!t]
	\centering
	\begin{overpic}[width=0.49\textwidth]{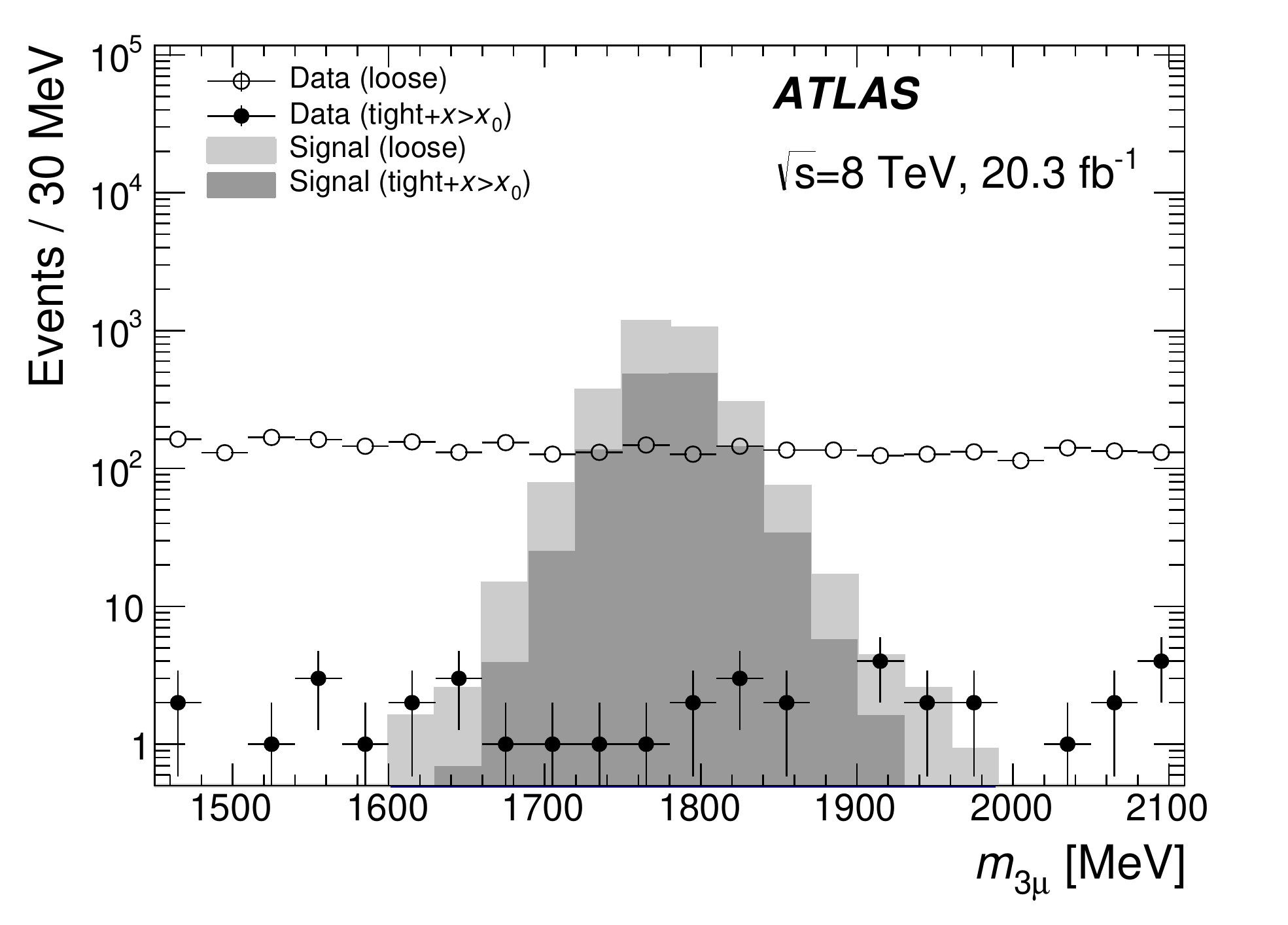}\put(25,45){\textrm{(a)}}\end{overpic}
	\begin{overpic}[width=0.49\textwidth]{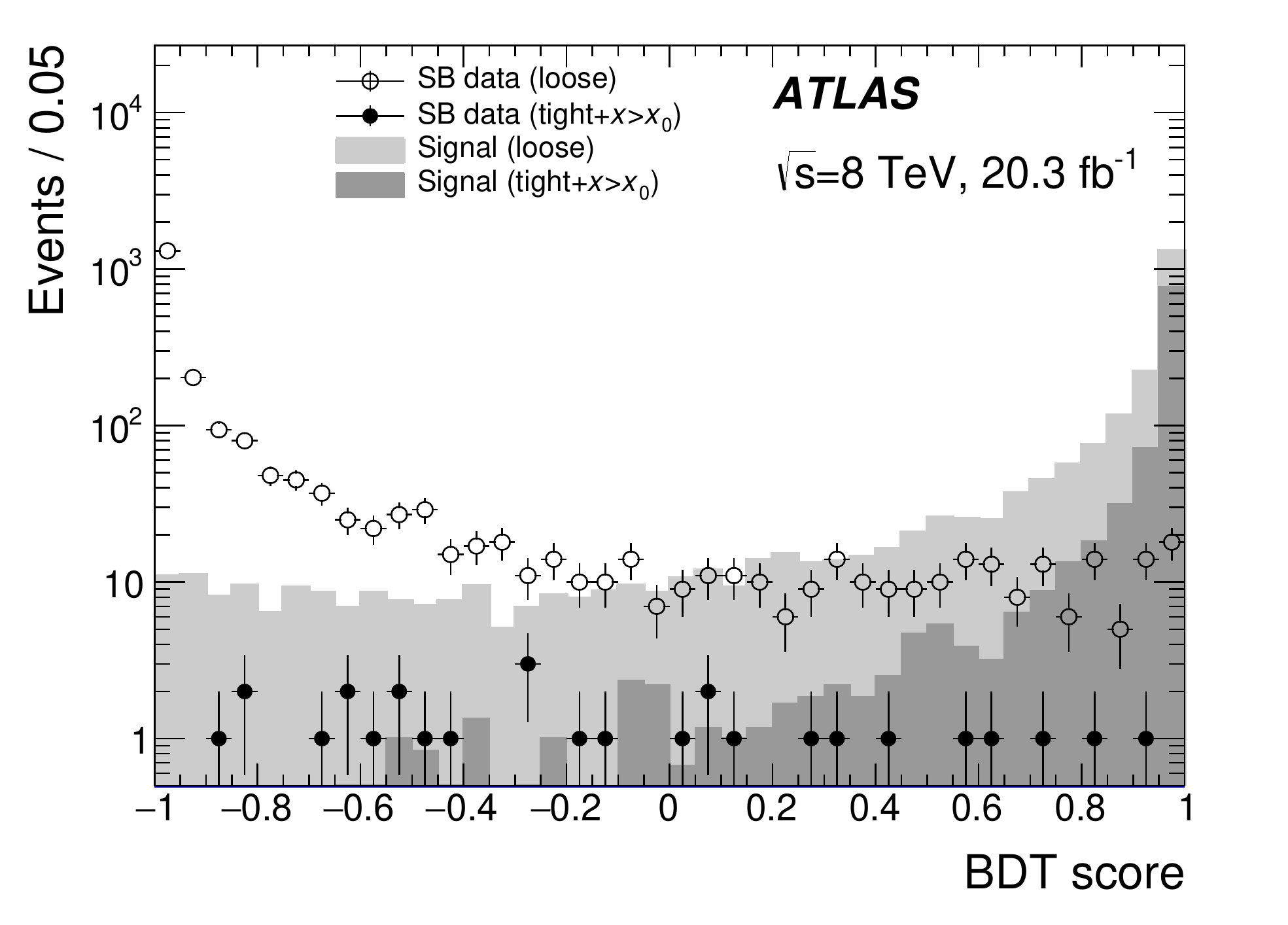}\put(25,45){\textrm{(b)}}\end{overpic}
	\caption{The three-muon mass distribution in (a) and the BDT score in (b). The BDT score distribution of the data is shown for the sideband region. The \loose data are shown as hollow circles, while the \loose signal MC events are shown as light solid grey area. The \tightxzero data are shown as the solid black circles, while the \tightxzero signal MC events are shown as the dark solid grey area. The area of the signal MC shapes is normalised to the area of the \loose data shapes and the relative normalisation difference between the \loose and the \tightxzero MC signal distributions prior to the normalisation is maintained. For illustration, the signal is not constrained to the SR.}
	\label{fig:mass_and_score}
\end{figure}

\begin{figure}[!t]
	\centering
	\begin{overpic}[width=0.43\textwidth]{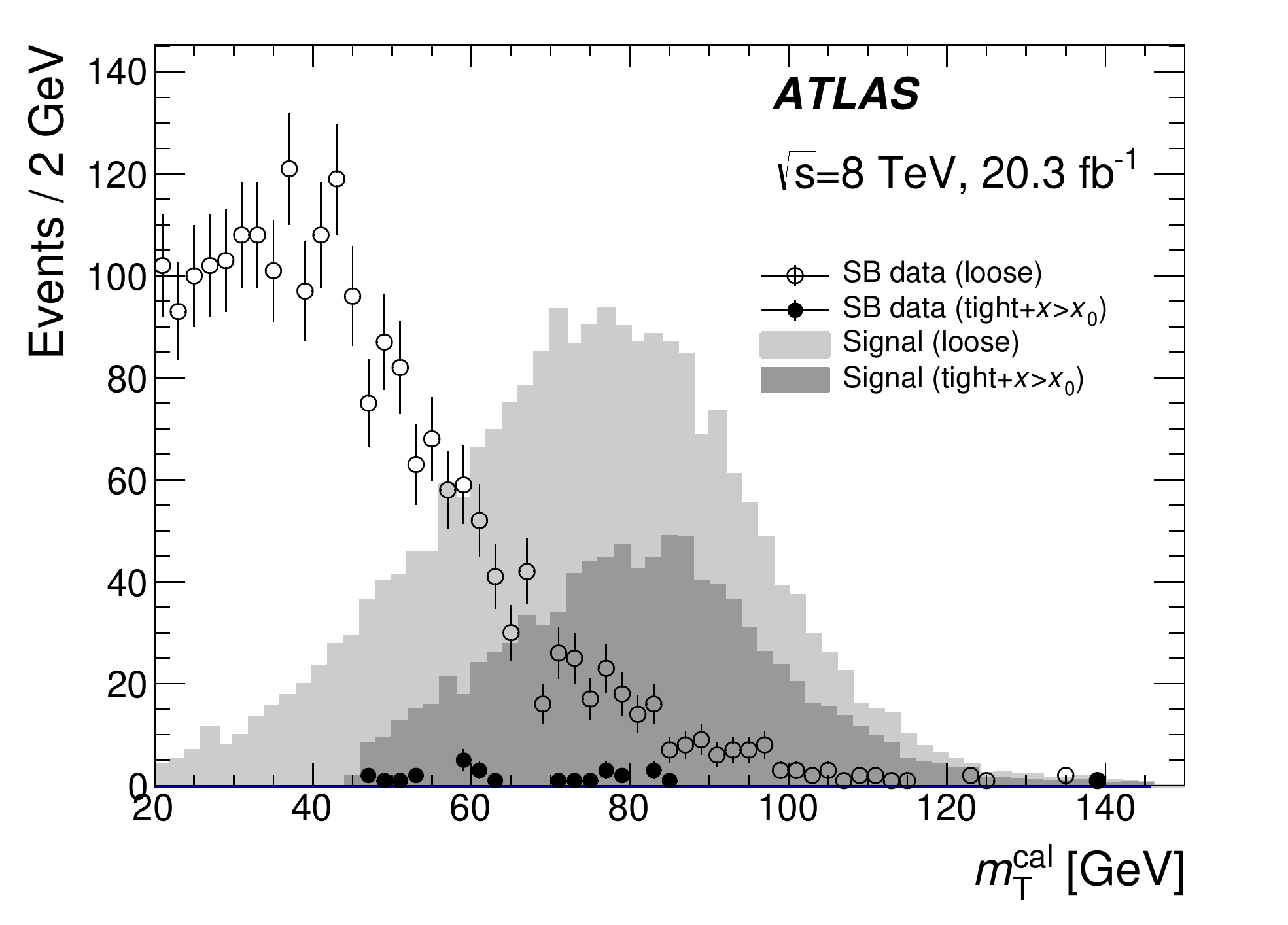}\put(75,25){\textrm{(a)}}\end{overpic}
	\begin{overpic}[width=0.43\textwidth]{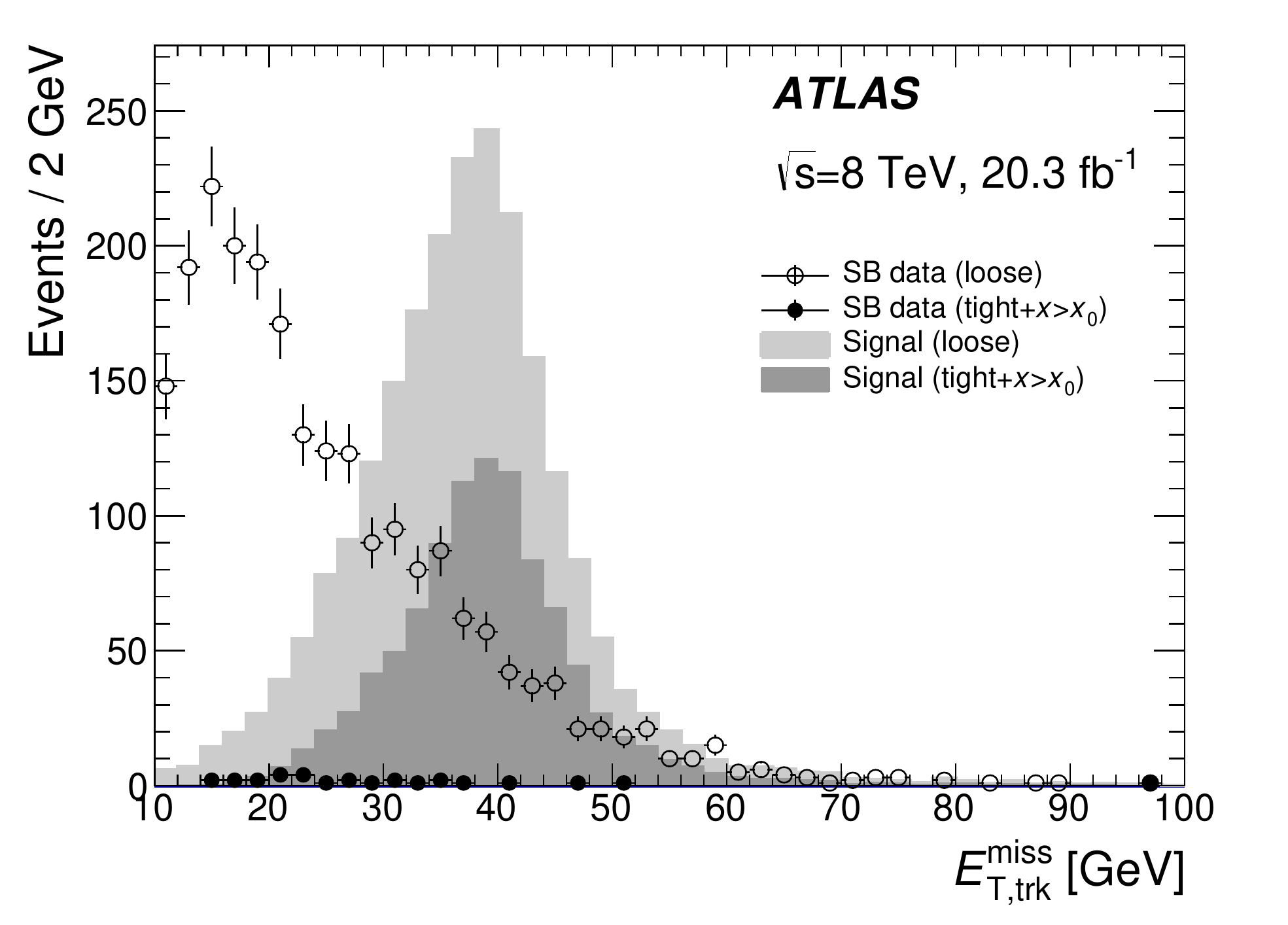}\put(75,25){\textrm{(b)}}\end{overpic}
	\begin{overpic}[width=0.43\textwidth]{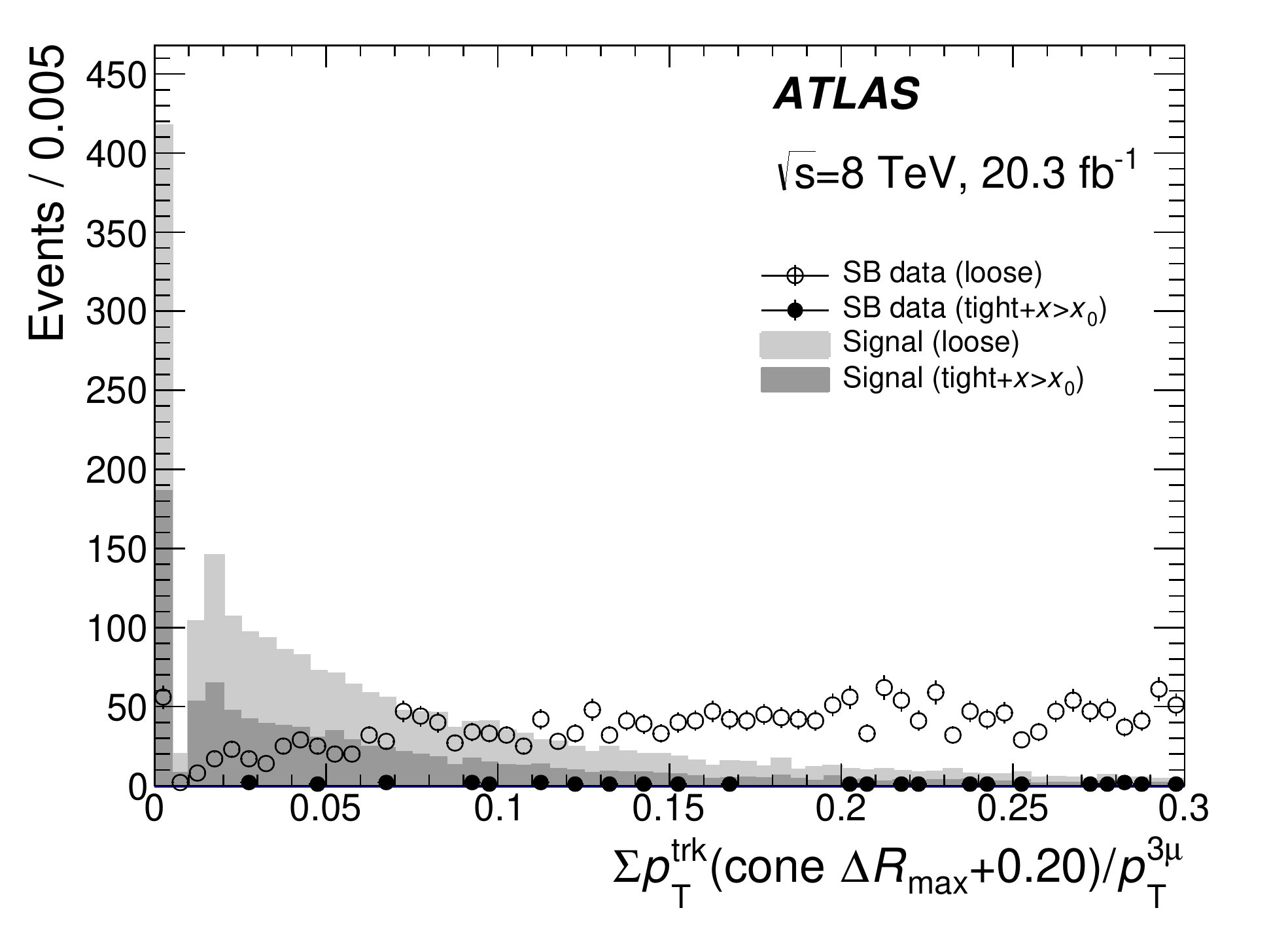}\put(75,25){\textrm{(c)}}\end{overpic}
	\begin{overpic}[width=0.43\textwidth]{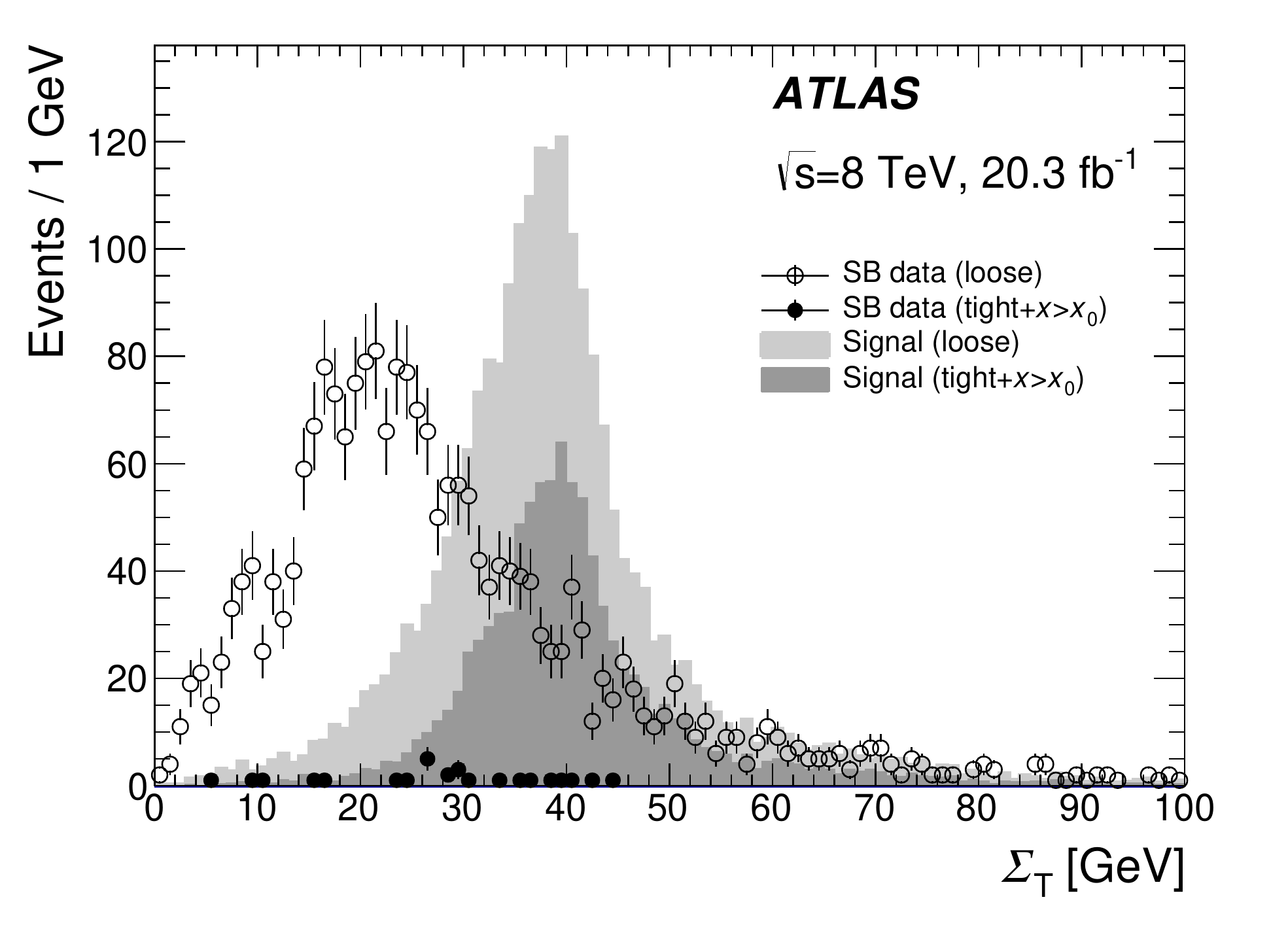}\put(75,25){\textrm{(d)}}\end{overpic}
	\begin{overpic}[width=0.43\textwidth]{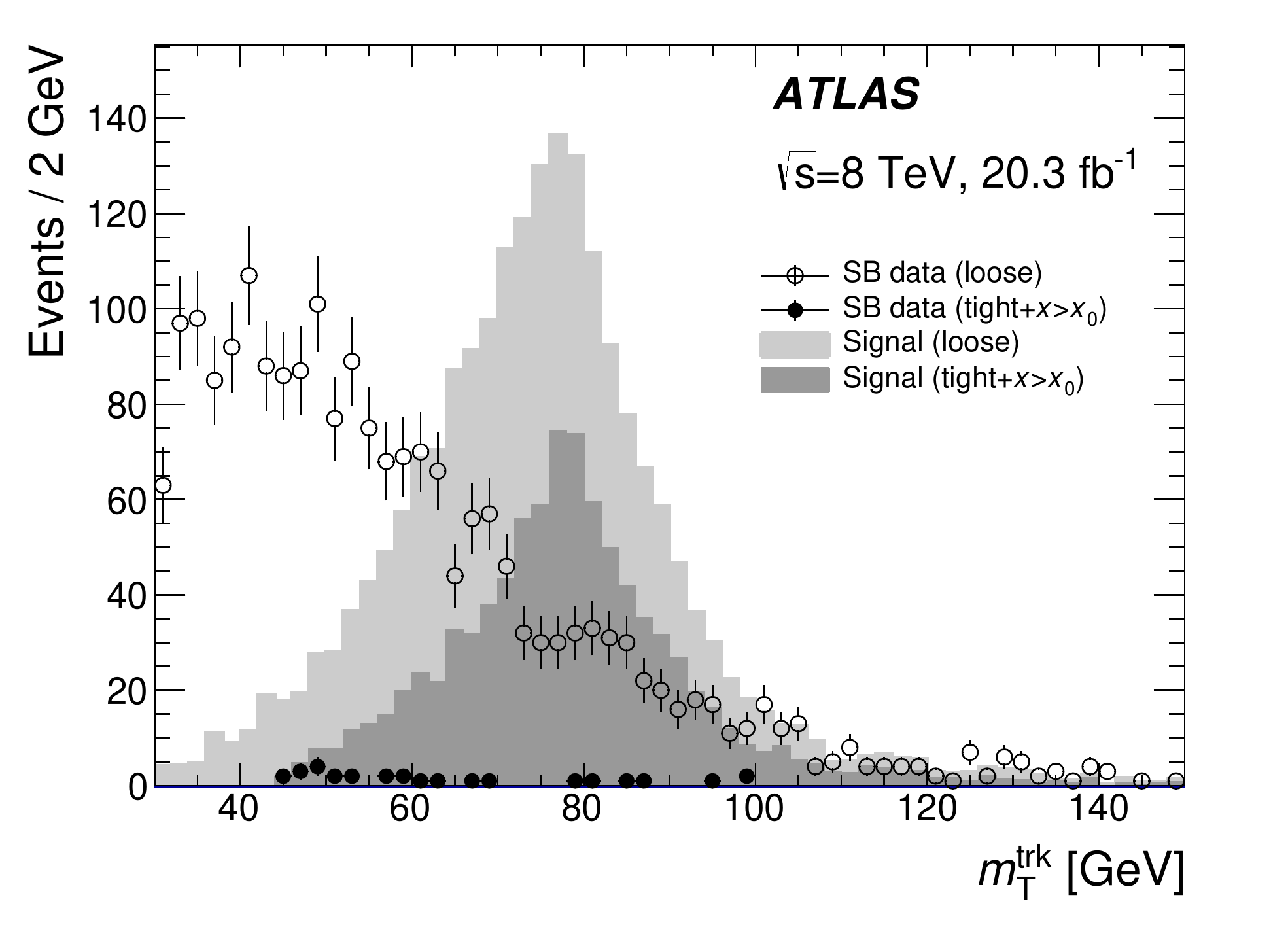}\put(75,25){\textrm{(e)}}\end{overpic}
	\begin{overpic}[width=0.43\textwidth]{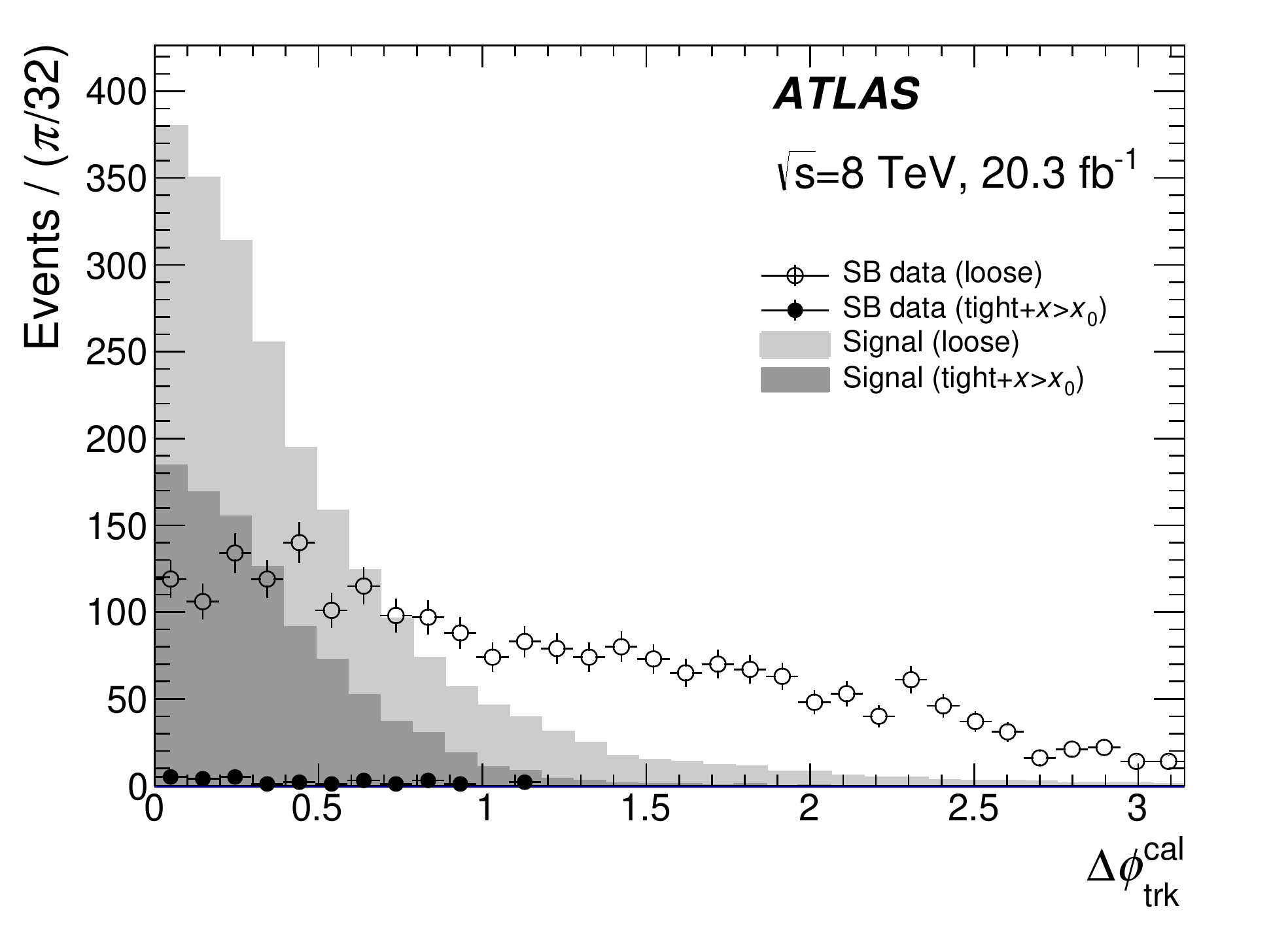}\put(75,25){\textrm{(f)}}\end{overpic}
	\begin{overpic}[width=0.43\textwidth]{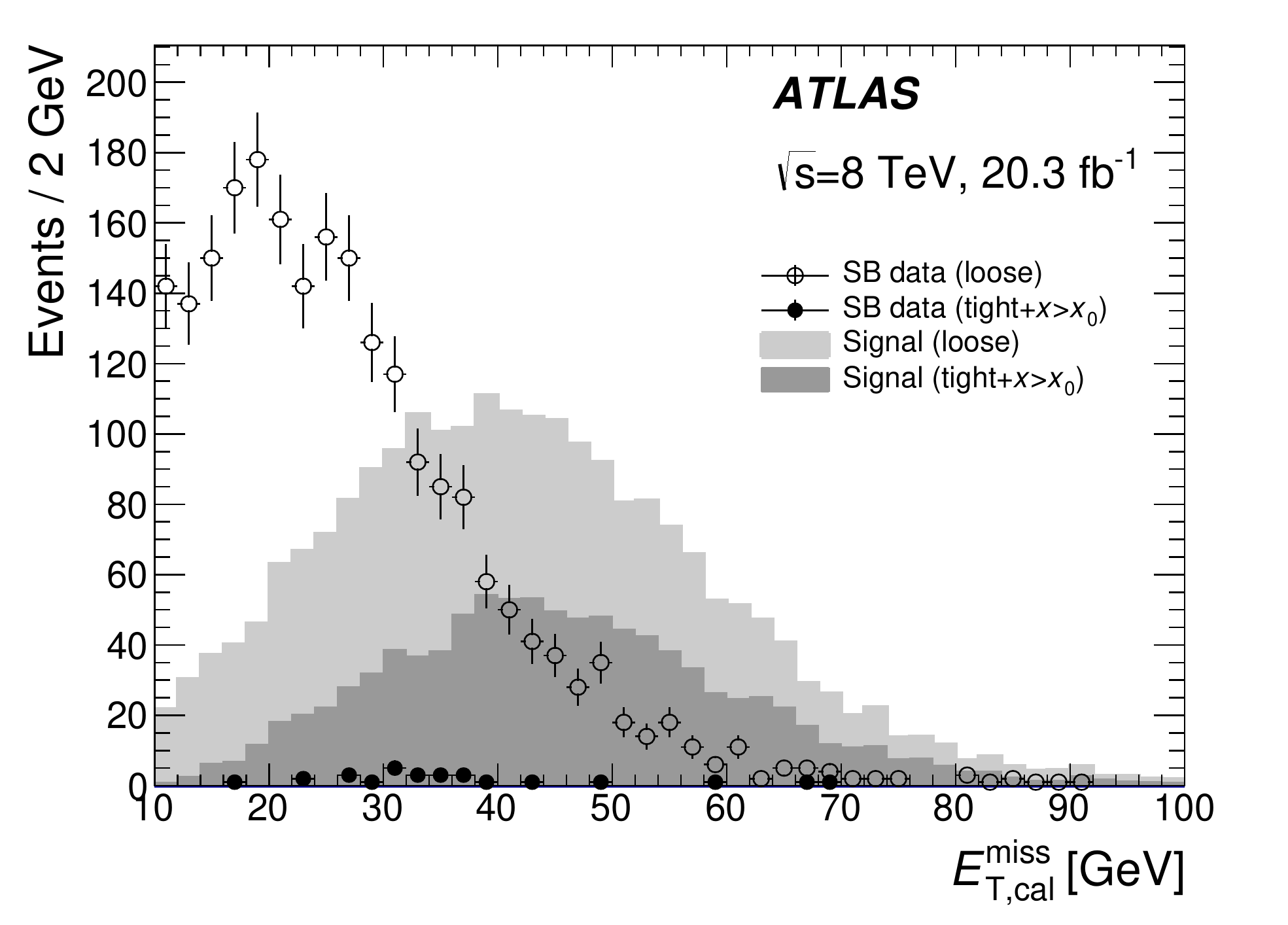}\put(75,25){\textrm{(g)}}\end{overpic}
	\begin{overpic}[width=0.43\textwidth]{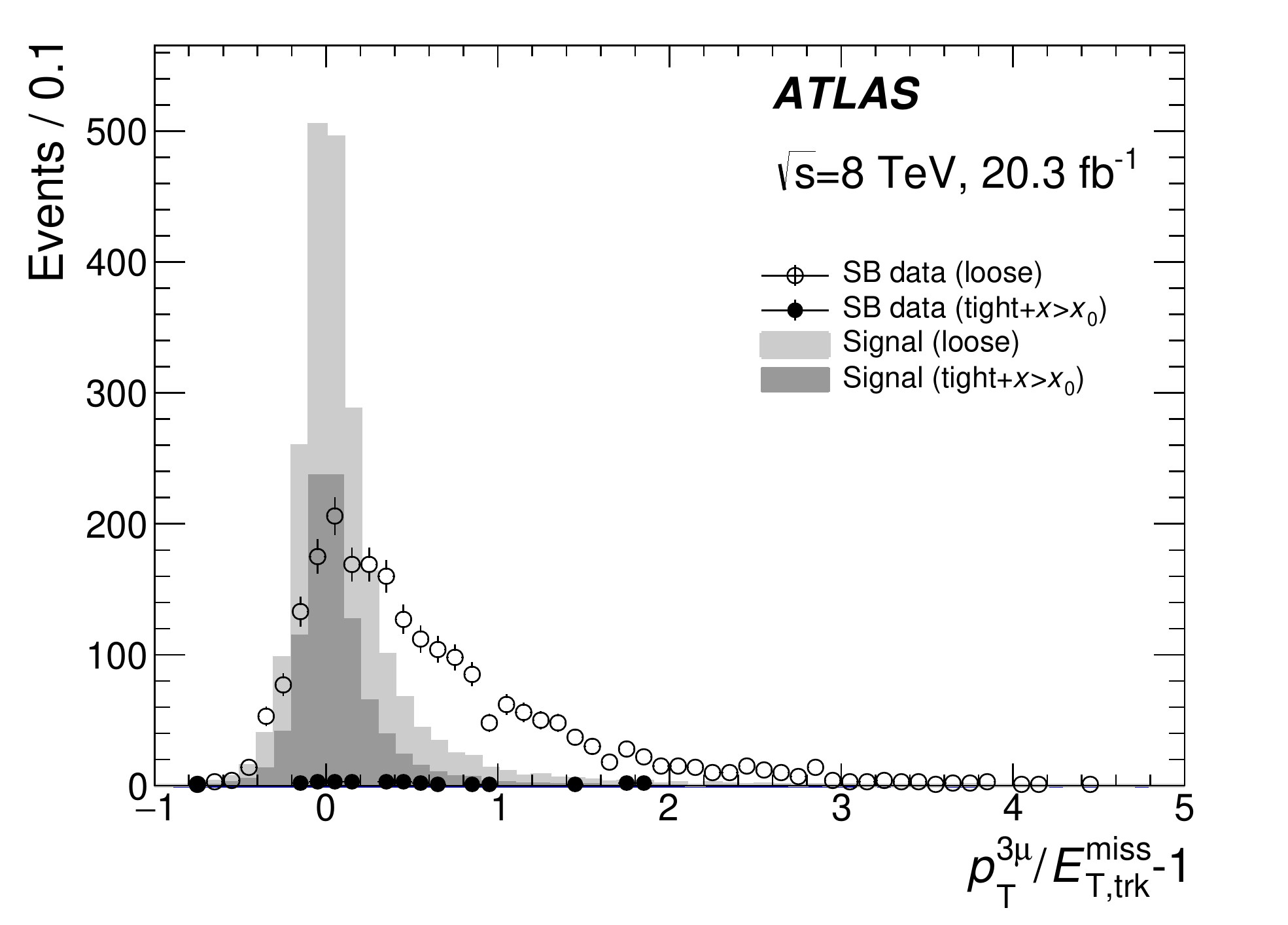}\put(75,25){\textrm{(h)}}\end{overpic}
	\caption{The BDT inputs ranked 1--8.  \mTcalo in (a), \trackmet in (b), \isoTwenty in (c), \HT in (d), \mTtrack in (e), \dphimets in (f), \calomet in (g) and $\pTthreebody/\tracketmis-1$ in (h). The \loose data in the sidebands are shown as hollow circles, while the \loose signal MC events are shown as light solid grey area. The \tightxzero data in the sidebands are shown as the solid black circles, while the \tightxzero signal MC events are shown as the dark solid grey area. The area of the signal MC shapes is normalised to the area of the \loose data shapes and the relative normalisation difference between the \loose and the \tightxzero MC signal distributions prior to the normalisation is maintained. For illustration, the signal is not constrained to the SR.}
	\label{fig:BDTinputs1}
\end{figure}

\begin{figure}[!t]
	\centering
	\begin{overpic}[width=0.43\textwidth]{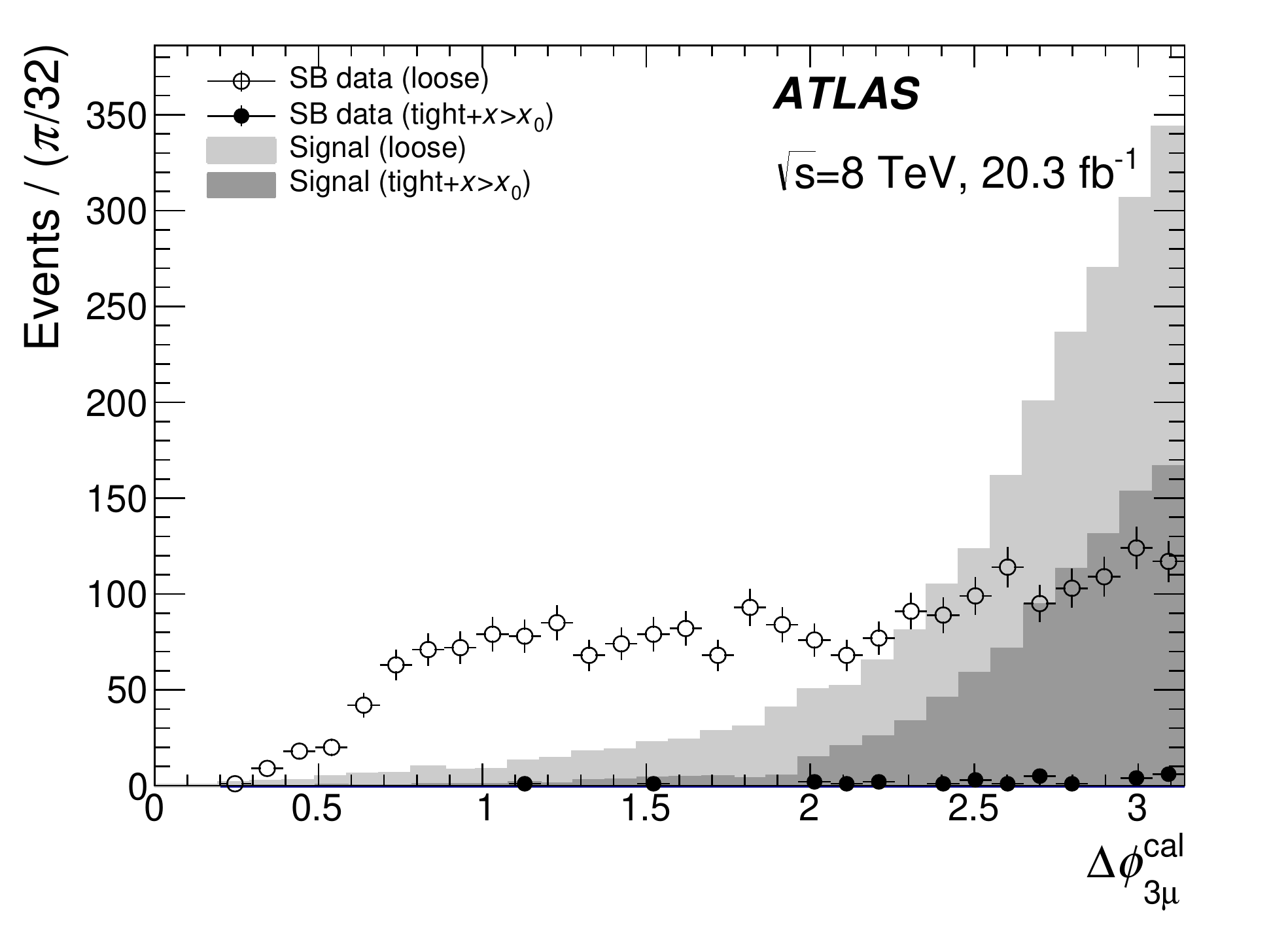}\put(20,25){\textrm{(a)}}\end{overpic}
	\begin{overpic}[width=0.43\textwidth]{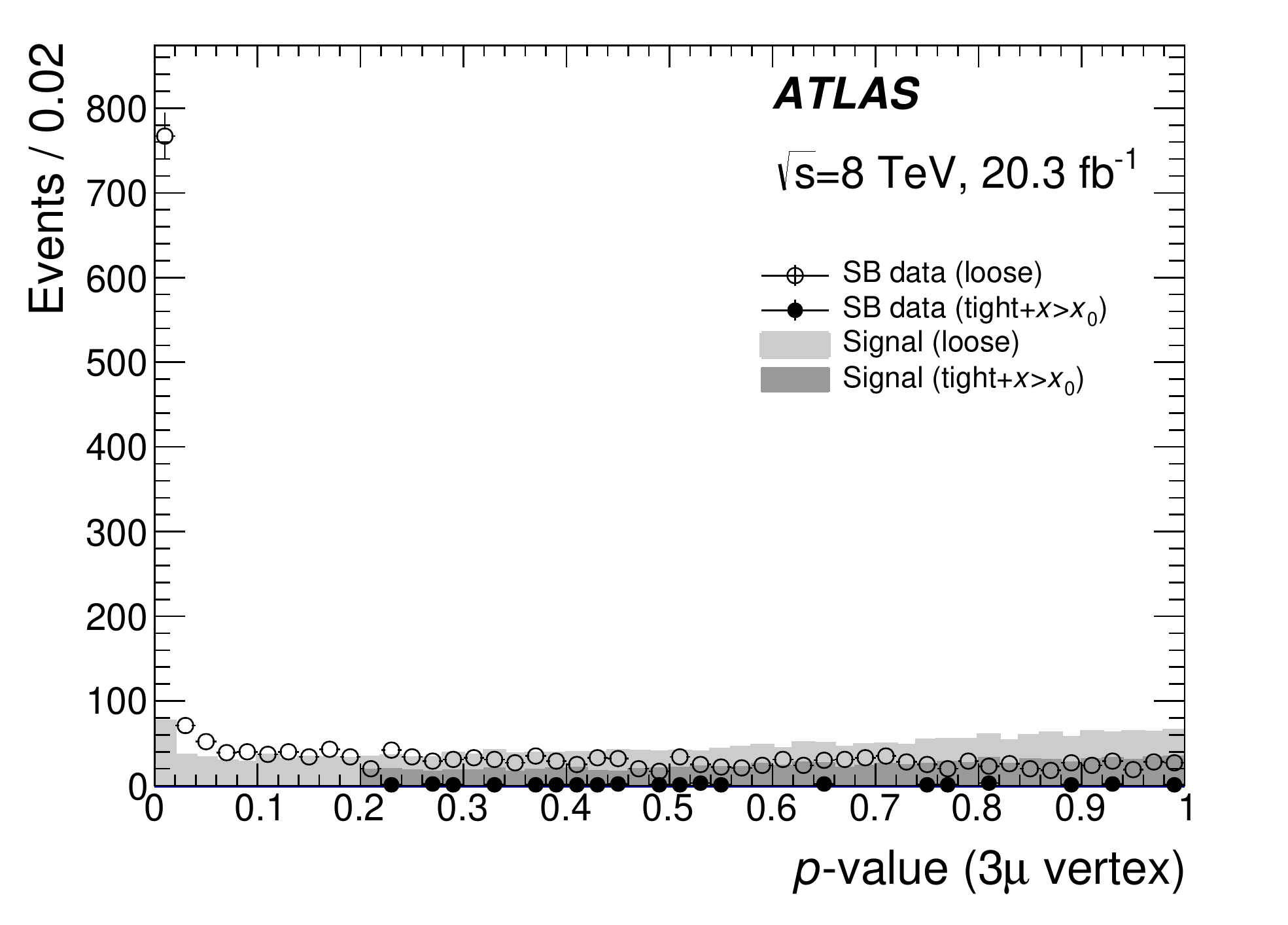}\put(20,25){\textrm{(b)}}\end{overpic}
	\begin{overpic}[width=0.43\textwidth]{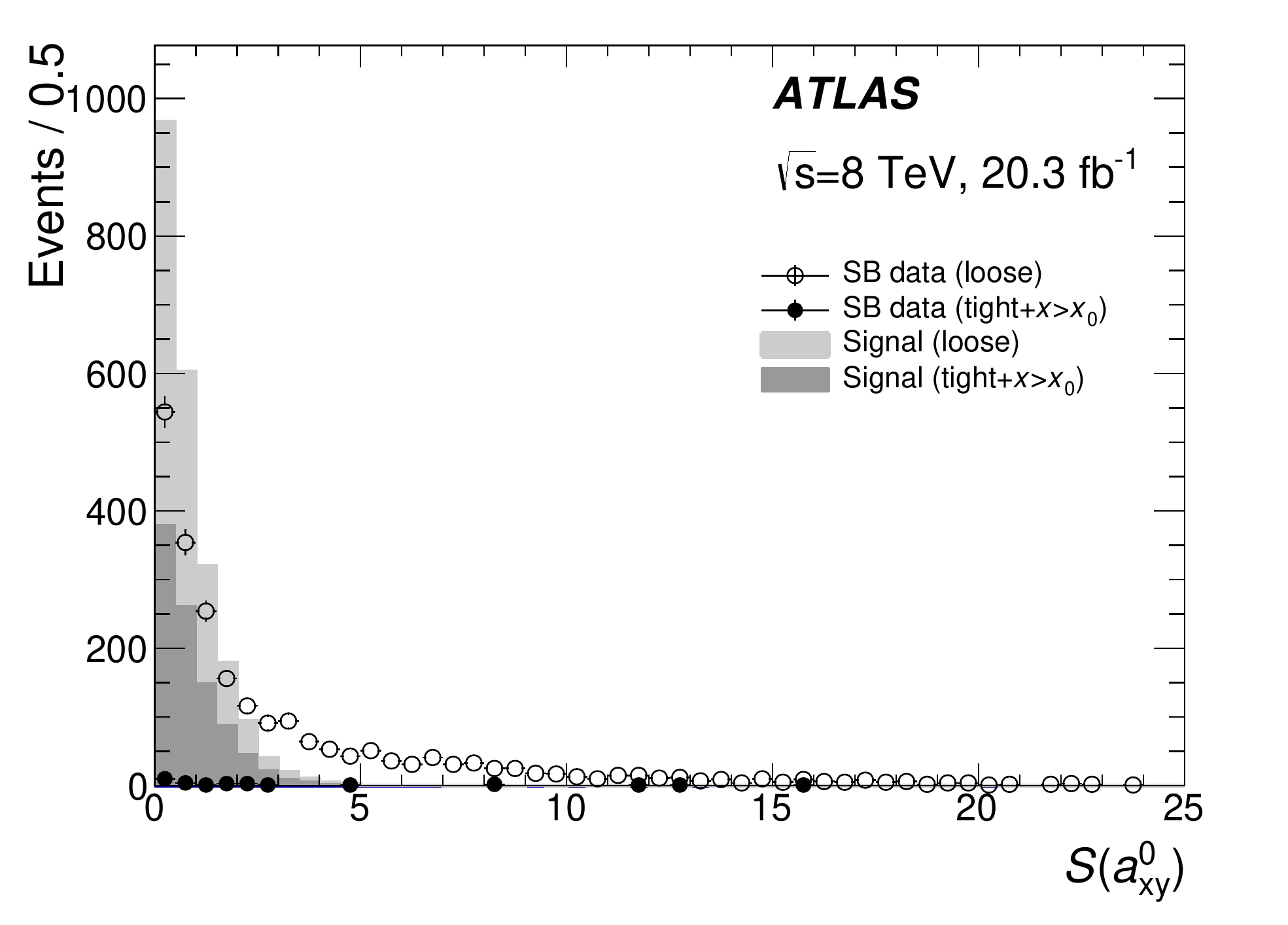}\put(20,25){\textrm{(c)}}\end{overpic}
	\begin{overpic}[width=0.43\textwidth]{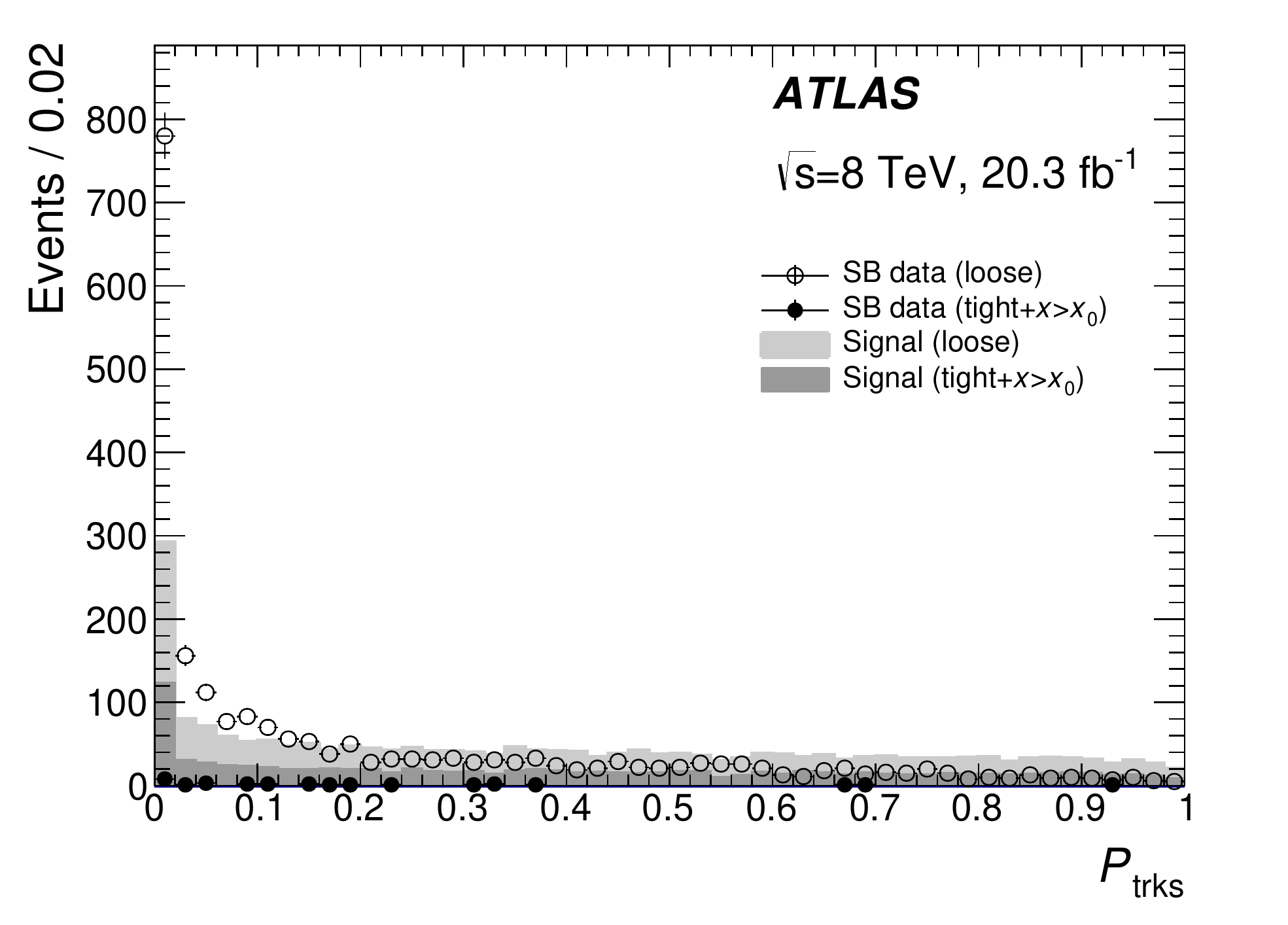}\put(20,25){\textrm{(d)}}\end{overpic}
	\begin{overpic}[width=0.43\textwidth]{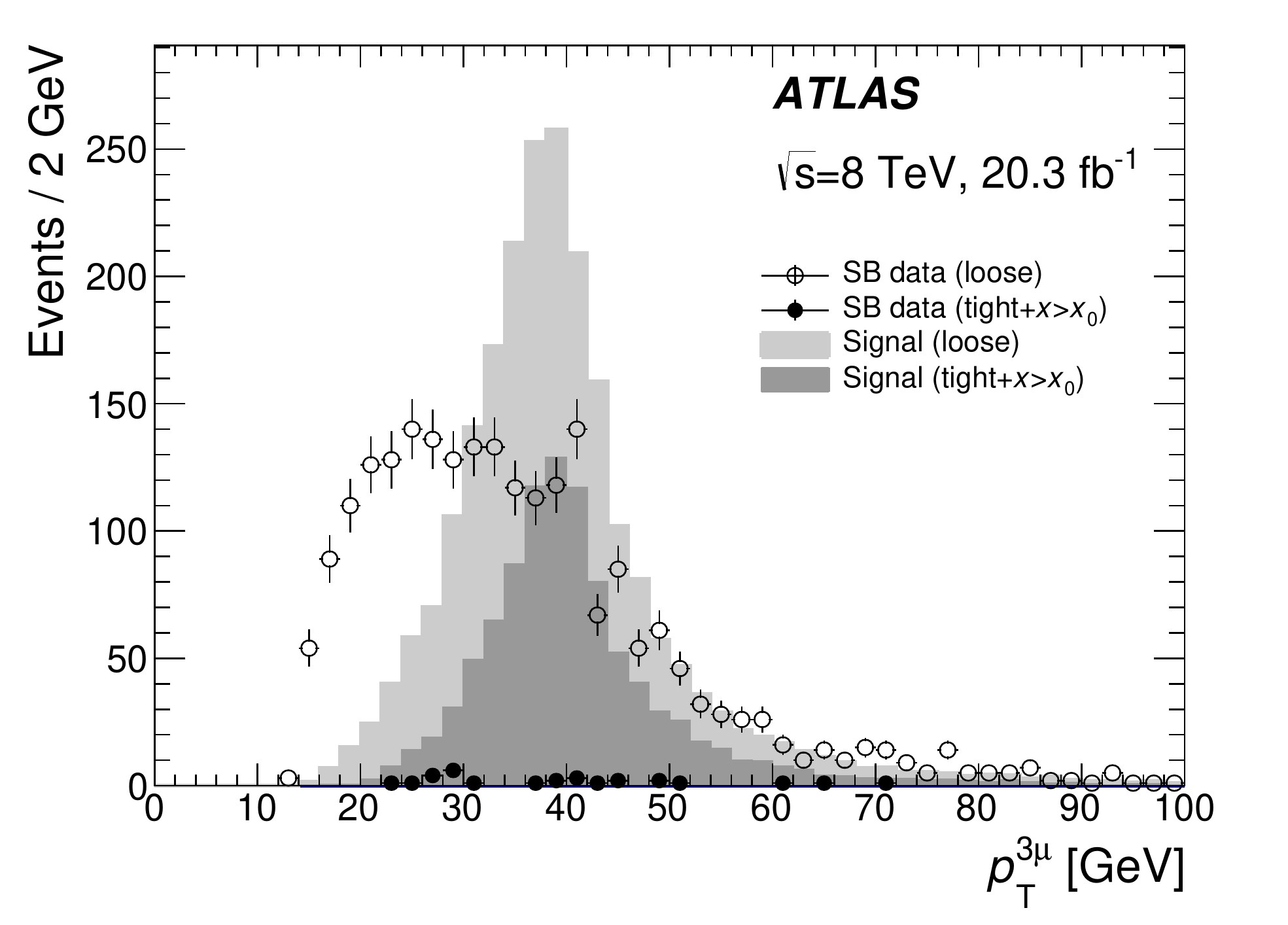}\put(75,25){\textrm{(e)}}\end{overpic}
	\begin{overpic}[width=0.43\textwidth]{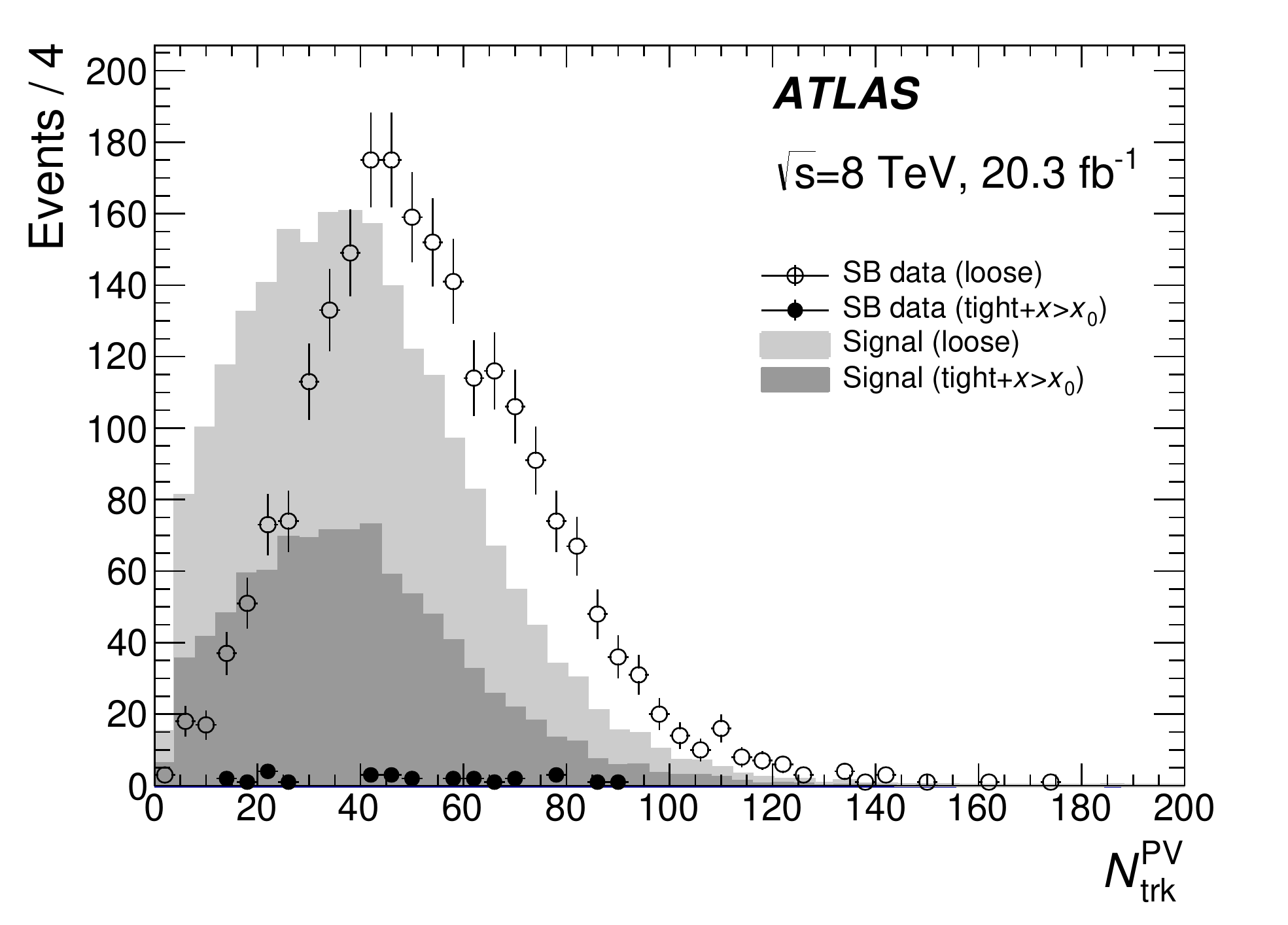}\put(75,25){\textrm{(f)}}\end{overpic}
	\begin{overpic}[width=0.43\textwidth]{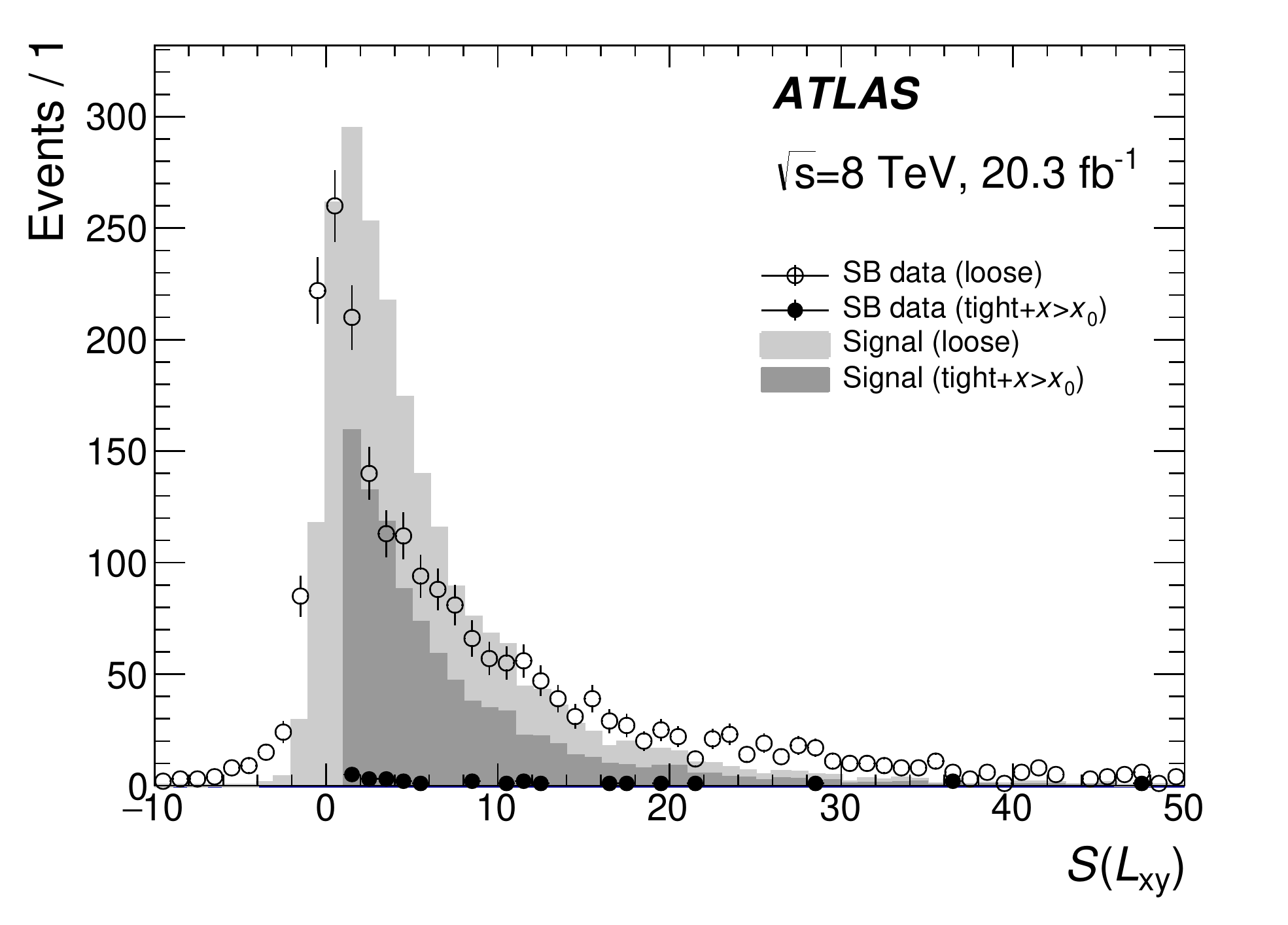}\put(75,25){\textrm{(g)}}\end{overpic}
	\begin{overpic}[width=0.43\textwidth]{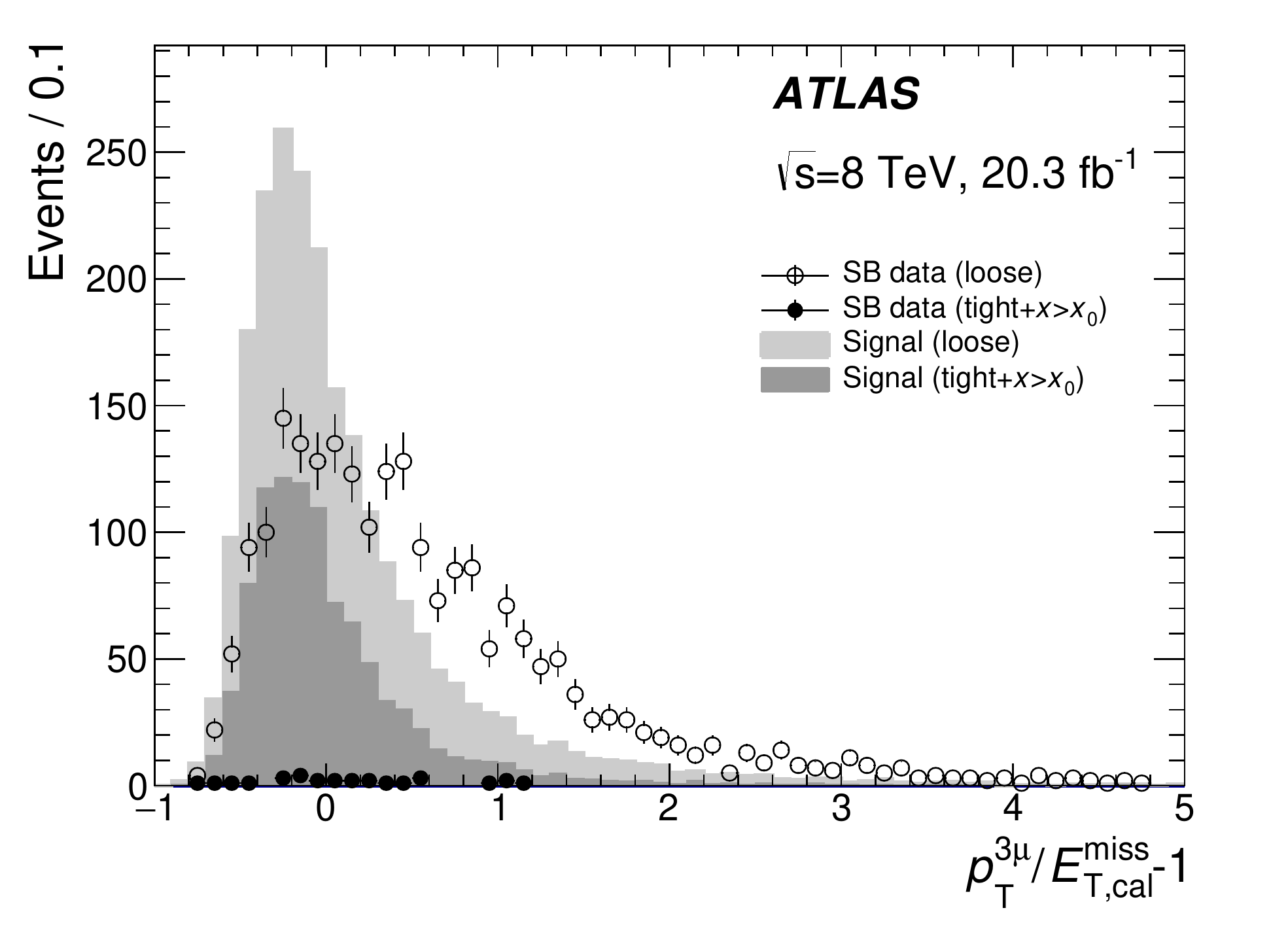}\put(75,25){\textrm{(h)}}\end{overpic}
	\caption{The BDT inputs ranked 9--16. \dphithreebodymetcalo in (a), \pvalue in (b), \Sazeroxy in (c) \ptrks in (d), \pTthreebody in (e), \pvntrk in (f), \SLxy in (g) and $\pTthreebody/\caloetmis-1$ in (h). The \loose data in the sidebands are shown as hollow circles, while the \loose signal MC events are shown as light solid grey area. The \tightxzero data in the sidebands are shown as the solid black circles, while the \tightxzero signal MC events are shown as the dark solid grey area. The area of the signal MC shapes is normalised to the area of the \loose data shapes and the relative normalisation difference between the \loose and the \tightxzero MC signal distributions prior to the normalisation is maintained. For illustration, the signal is not constrained to the SR.}
	\label{fig:BDTinputs2}
\end{figure}

\begin{figure}[!t]
	\centering
	\begin{overpic}[width=0.32\textwidth]{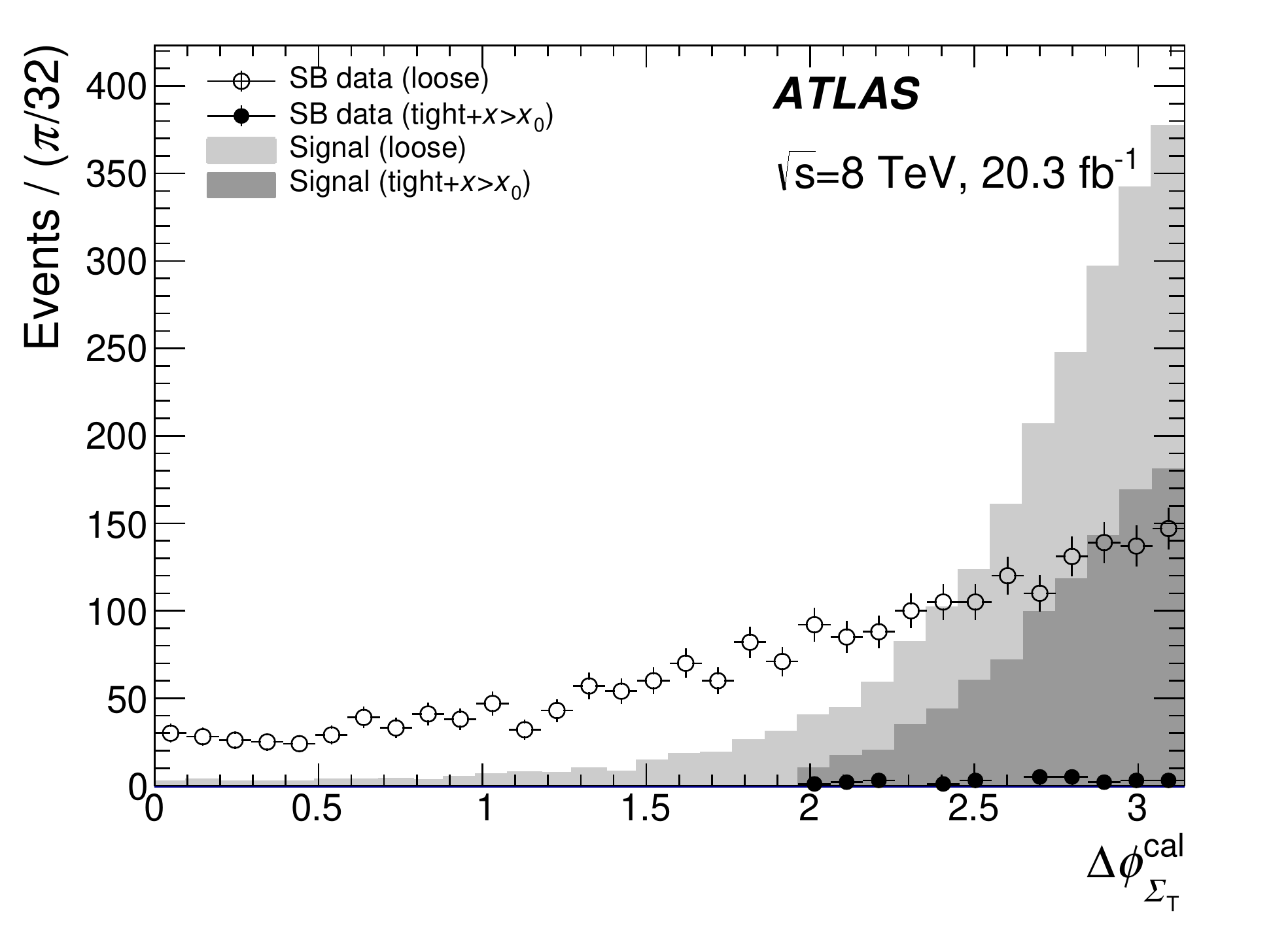}\put(25,25){\textrm{(a)}}\end{overpic}
	\begin{overpic}[width=0.32\textwidth]{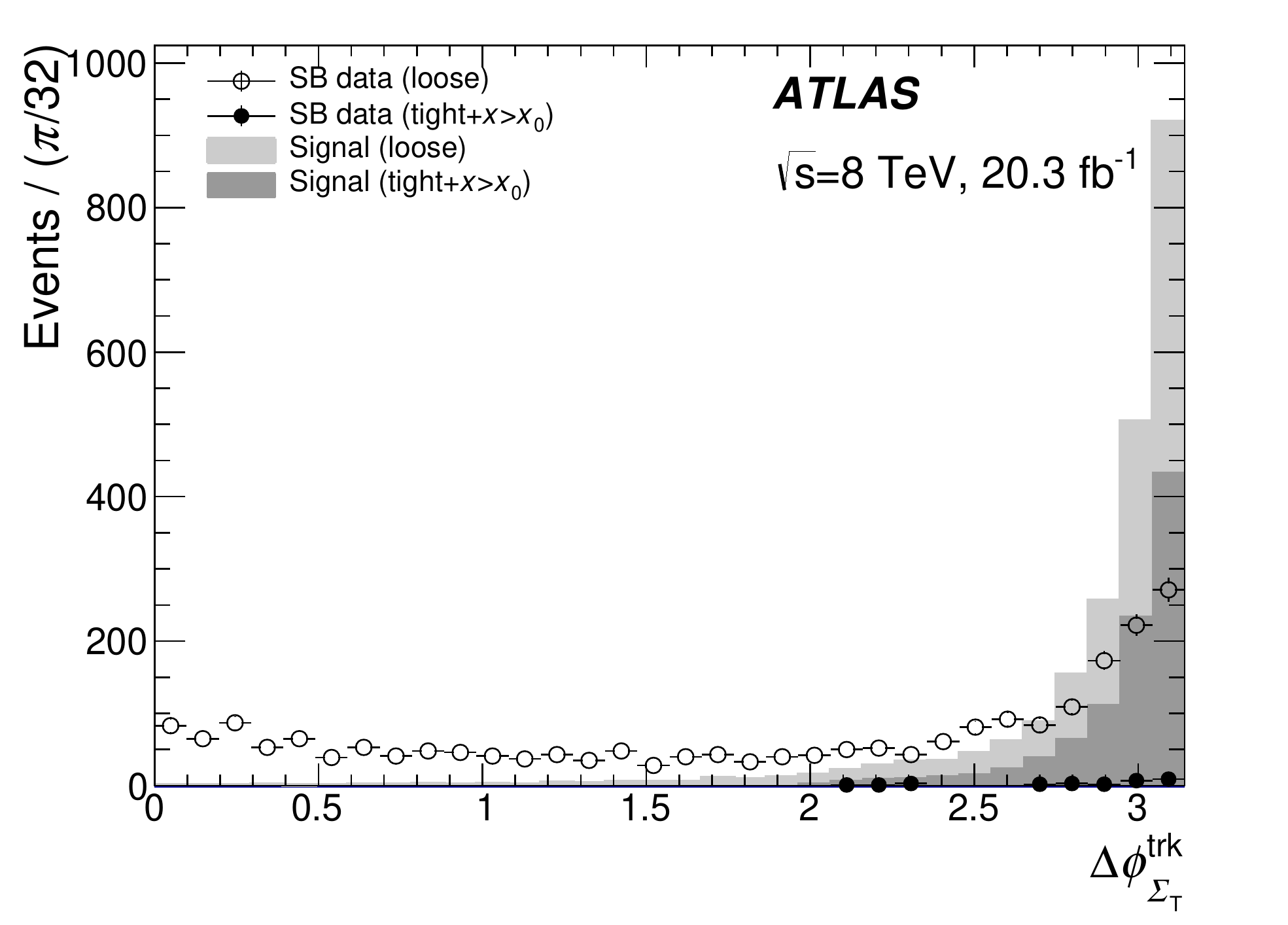}\put(25,25){\textrm{(b)}}\end{overpic}
	\begin{overpic}[width=0.32\textwidth]{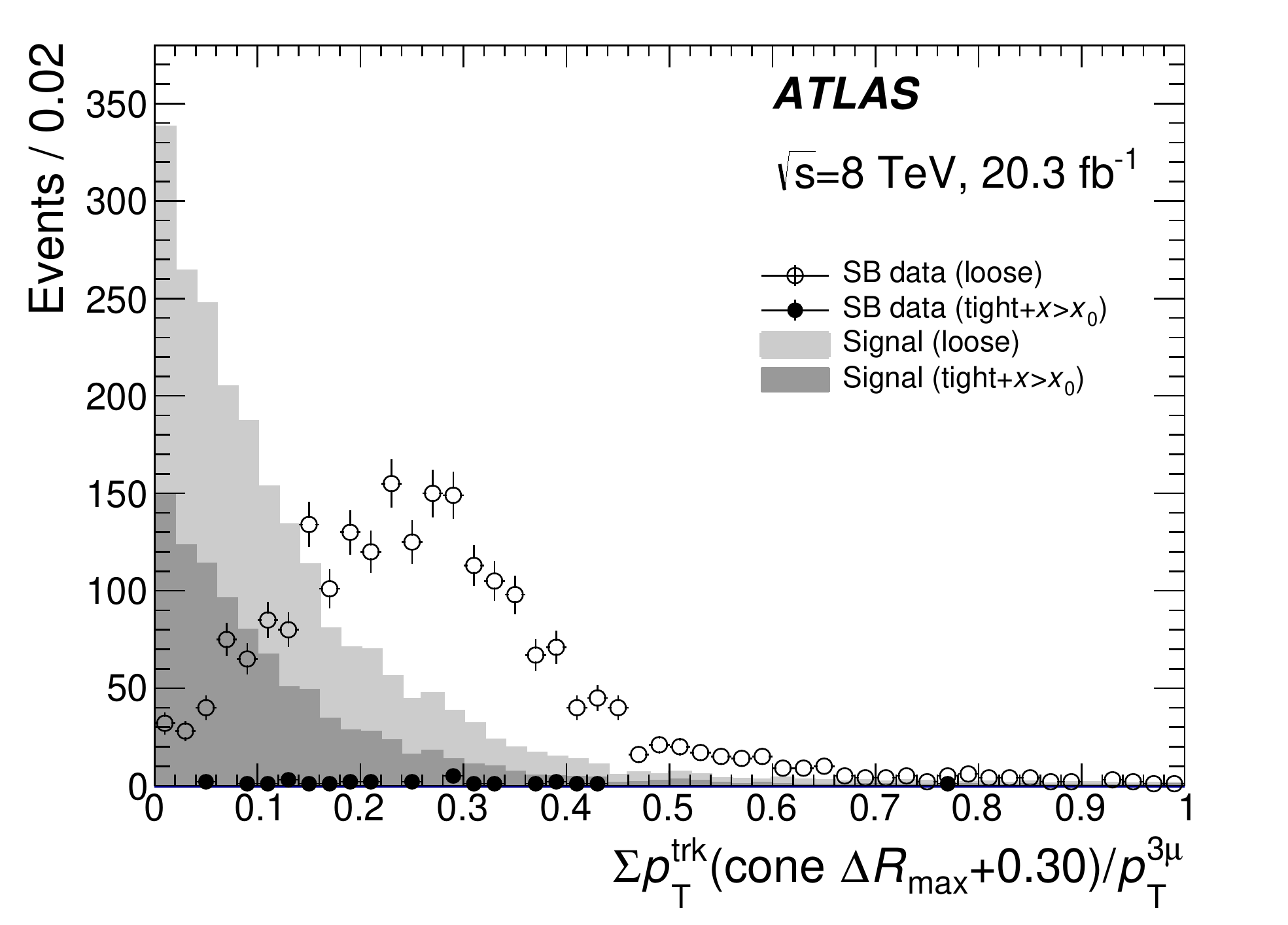}\put(75,25){\textrm{(c)}}\end{overpic}
	\begin{overpic}[width=0.32\textwidth]{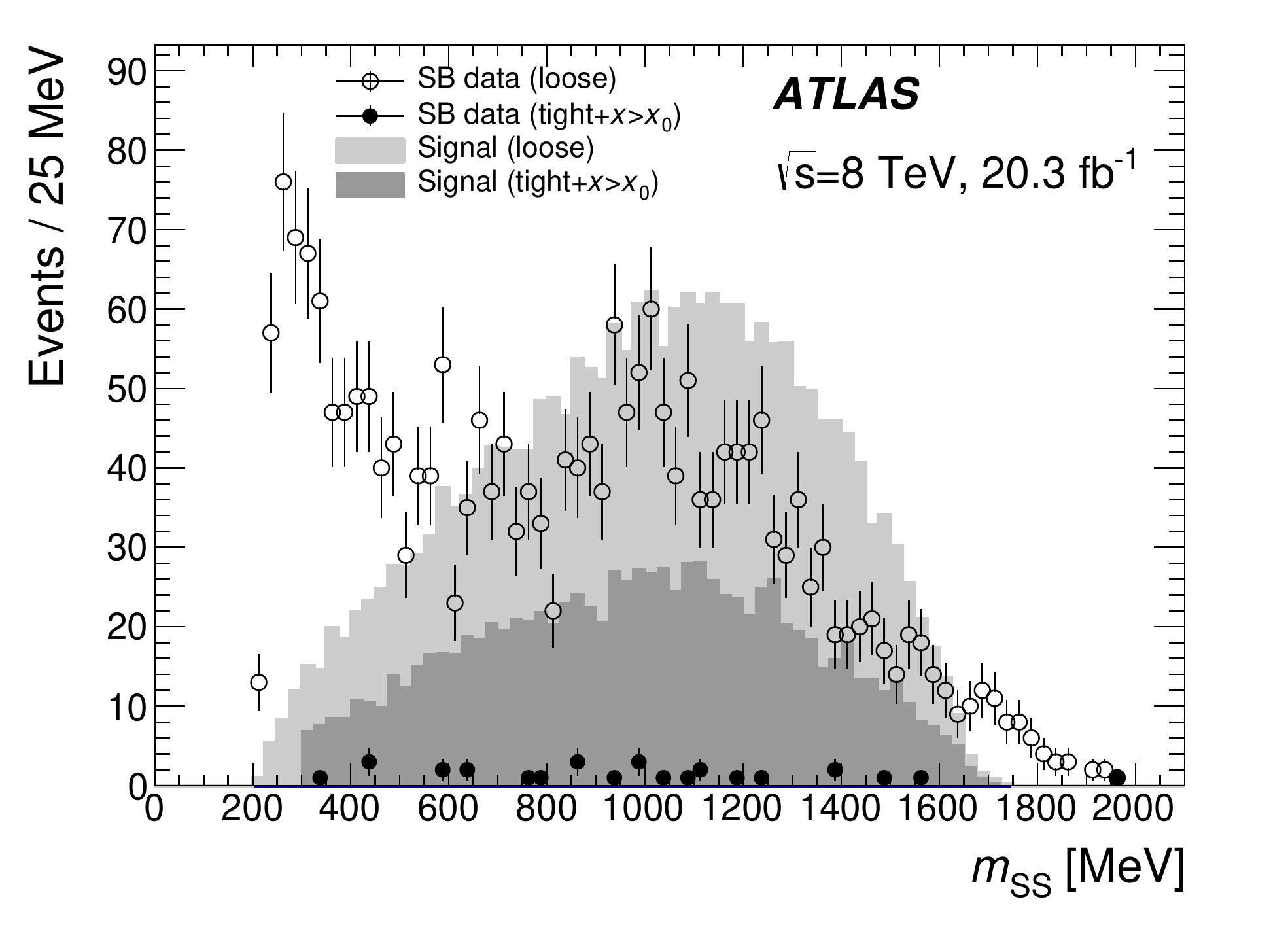}\put(75,25){\textrm{(d)}}\end{overpic}
	\begin{overpic}[width=0.32\textwidth]{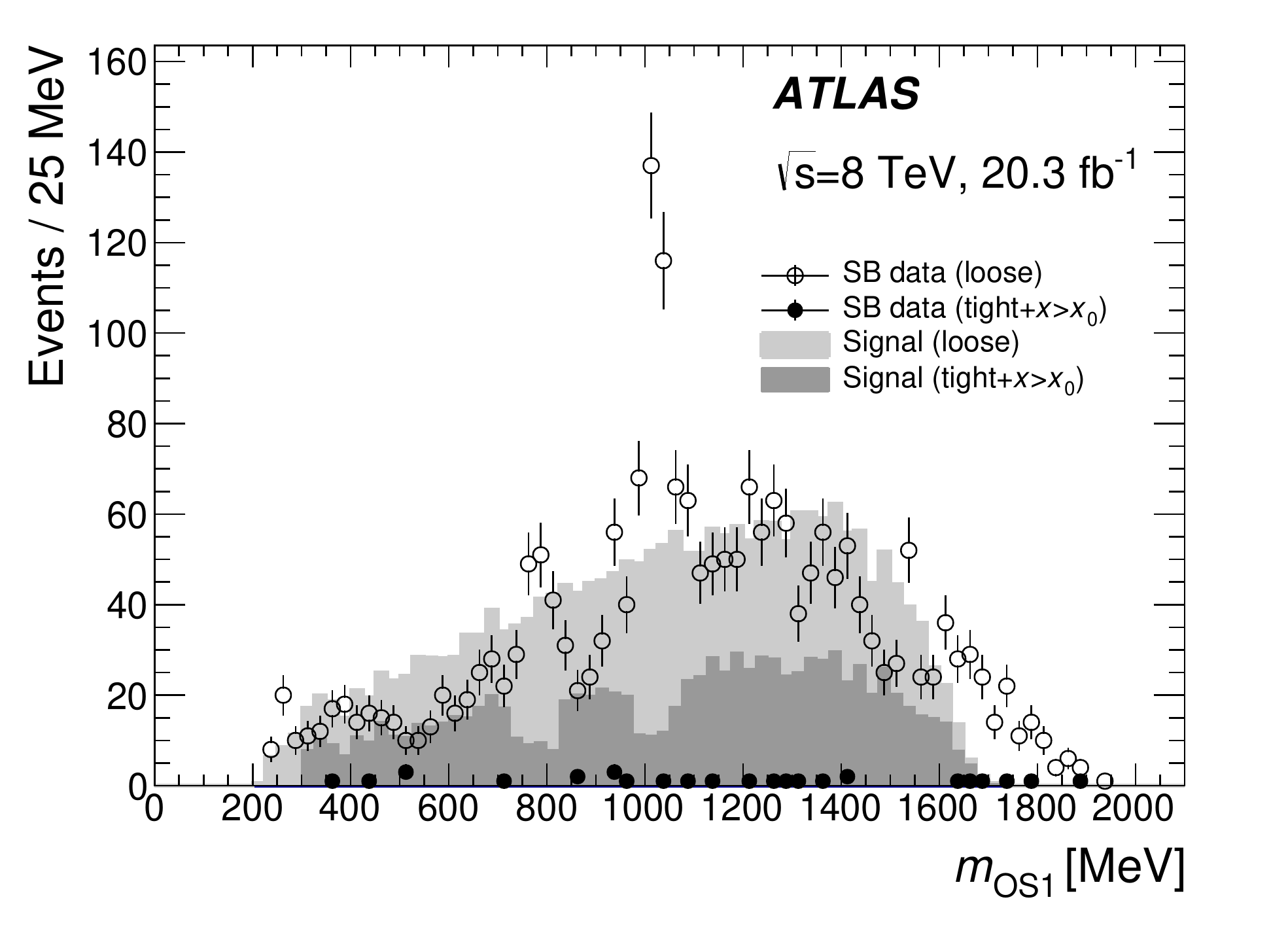}\put(75,25){\textrm{(e)}}\end{overpic}
	\begin{overpic}[width=0.32\textwidth]{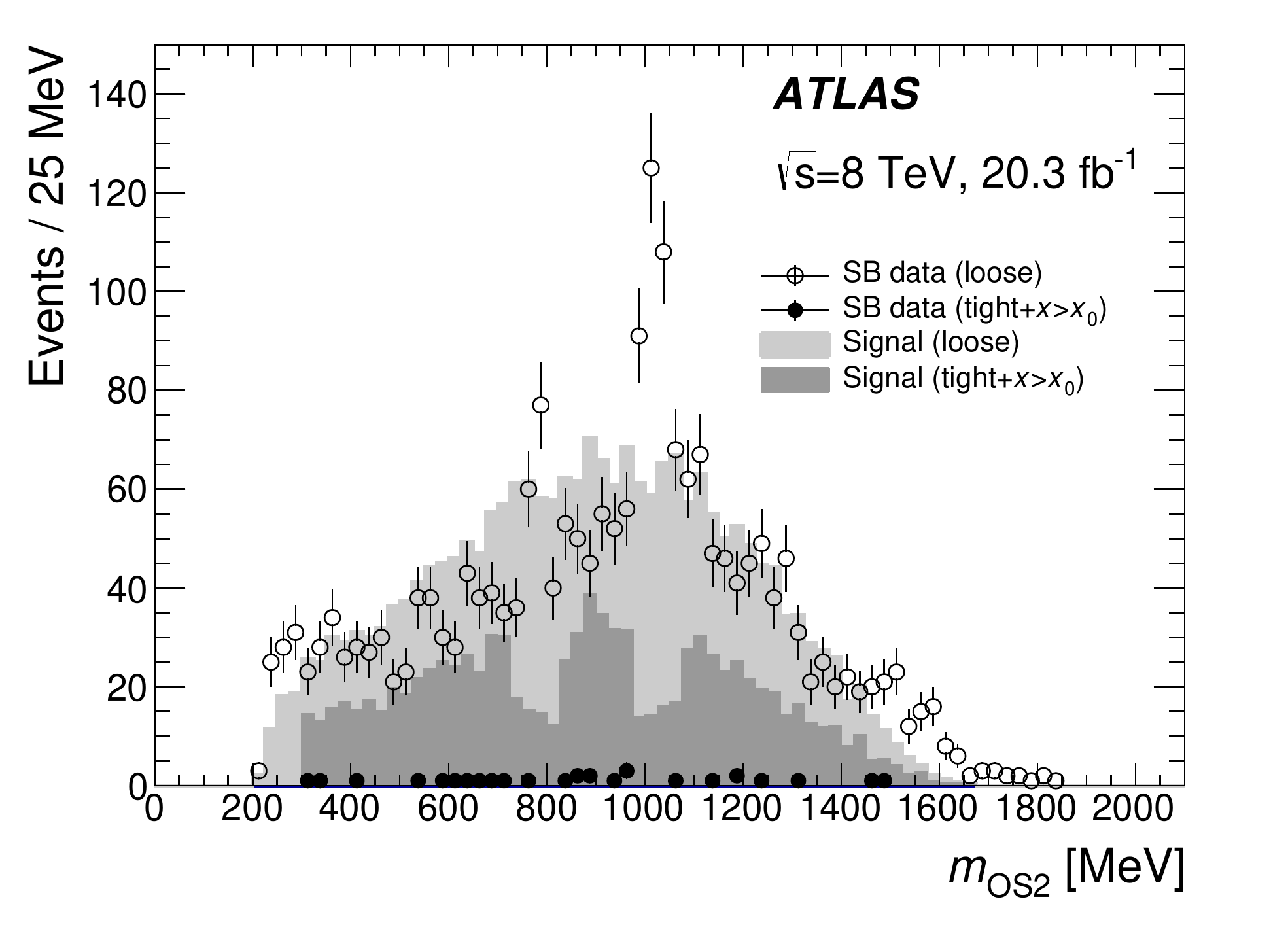}\put(75,25){\textrm{(f)}}\end{overpic}
	\caption{The complementary variables used in the \loose or \tight selection but not as inputs for the BDT. \dphiHmetcalo in (a), \dphiHmettrack in (b), \isoThirty in (c), \mSS in (d), \mOSa in (e) and \mOSb in (f). The \loose data in the sidebands are shown as hollow circles, while the \loose signal MC events are shown as light solid grey area. The \tightxzero data in the sidebands are shown as the solid black circles, while the \tightxzero signal MC events are shown as the dark solid grey area. The area of the signal MC shapes is normalised to the area of the \loose data shapes and the relative normalisation difference between the \loose and the \tightxzero MC signal distributions prior to the normalisation is maintained. For illustration, the signal is not constrained to the SR.}
	\label{fig:other}
\end{figure}

\subsection{Background estimation}
\label{sec:background_estimation}
The events passing the \tightxzero selection are used to estimate the expected number of background events in the signal region for higher cuts on $x$ as described below.

The signal MC and sideband data BDT responses are shown in Fig.~\ref{fig:BDTresponse} after the \tightxzero selection.
The distinct shapes illustrate the power of the method in separating the background from the signal.
The analytical function also shown in Fig.~\ref{fig:BDTresponse} is a result of a fit to the sideband data, excluding the blinded region, using an unbinned maximum-likelihood estimator.
The fit function used is $a_0 + a_1(x+1)^{a_2}+a_3(x+1)^{a_4}$, where $a_i$ are the free fit parameters.
The parameter $a_2$ is required to be negative while the other are required to be non-negative.
This function can exhibit rising behaviour at both ends of the $x$ distribution ($x\to\pm 1$) and it is used to scale the quantities measured in \xzero to the corresponding quantities in \xone as explained below.
\begin{figure}[!t]
\centering\includegraphics[width=0.8\textwidth]{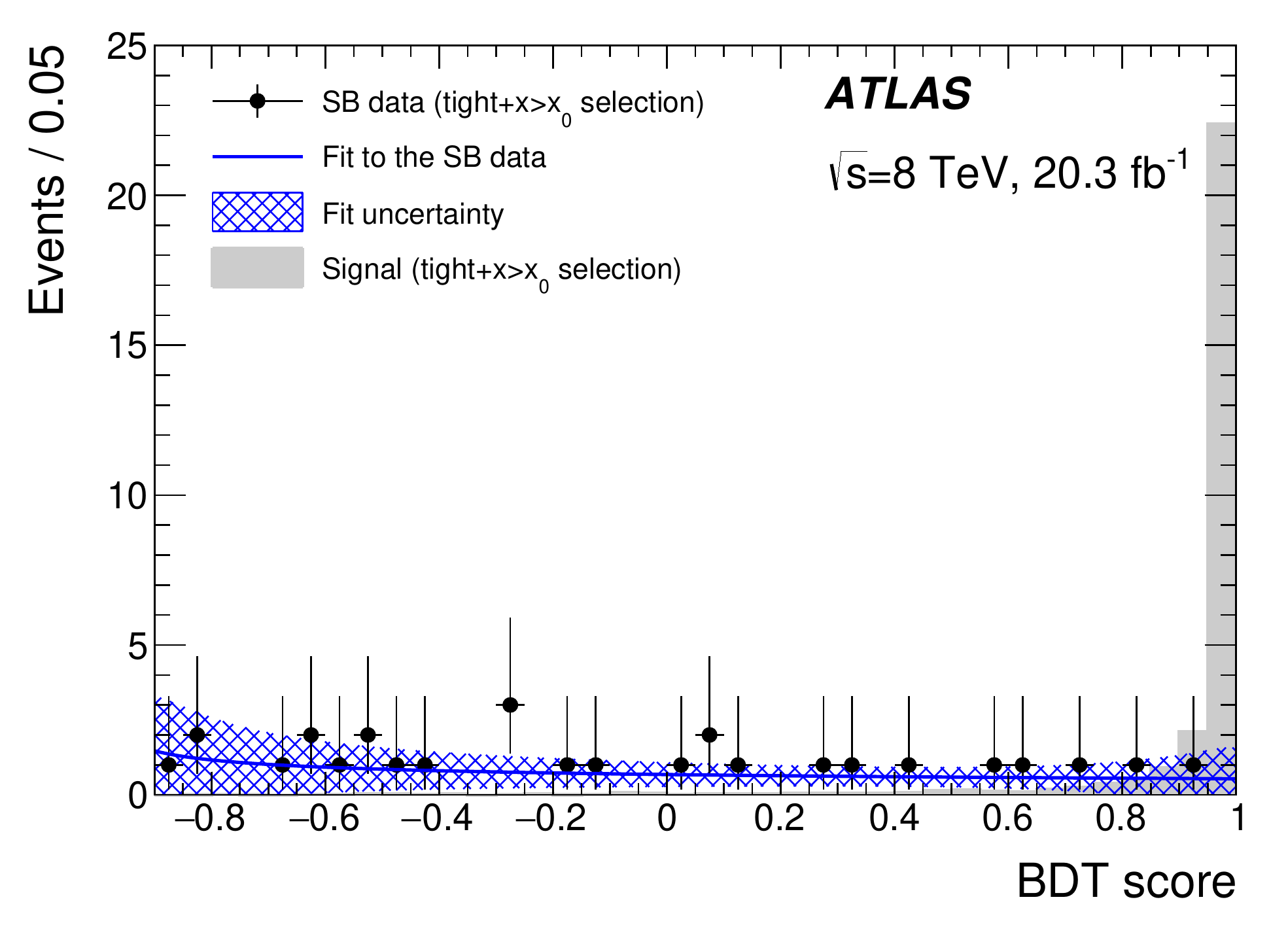}
\caption{The distribution of the BDT score of the data in the sideband region (SB) for the \tightxzero selection. The line shows the result of a fit to the BDT score distribution, while the hatched area shows the uncertainty in the fit due to the SB range definition, the $x_0$ cut location and the fit function choice. The solid grey area shows the signal shape (obtained from MC simulation), normalised to the area of the data for the \tightxzero selection. For illustration, the signal is not constrained to the SR.}
\label{fig:BDTresponse}
\end{figure}

The three-muon mass distribution of the \tightxzero data is fit simultaneously in the two sidebands to a second-order polynomial in \mthreebody while excluding the blinded region.
This is also done with an unbinned maximum-likelihood estimator.
The integral of the resulting fit function in the signal region gives the expected number of background events, \NbkgZero in the signal region before applying the final $x_1$ cut.
The statistical uncertainty of \NbkgZero is calculated by scaling the statistical error in the number of events in the sidebands, according to the ratio of analytical integrals in the signal region and sidebands.
Figure~\ref{fig:SB} shows the three-muon mass distribution in the sidebands for the \tightxzero selection as black points together with the fit result.
The signal is also shown for reference, scaled up arbitrarily to match the scale of the data.
\begin{figure}[!t]
\centering\includegraphics[width=0.8\textwidth]{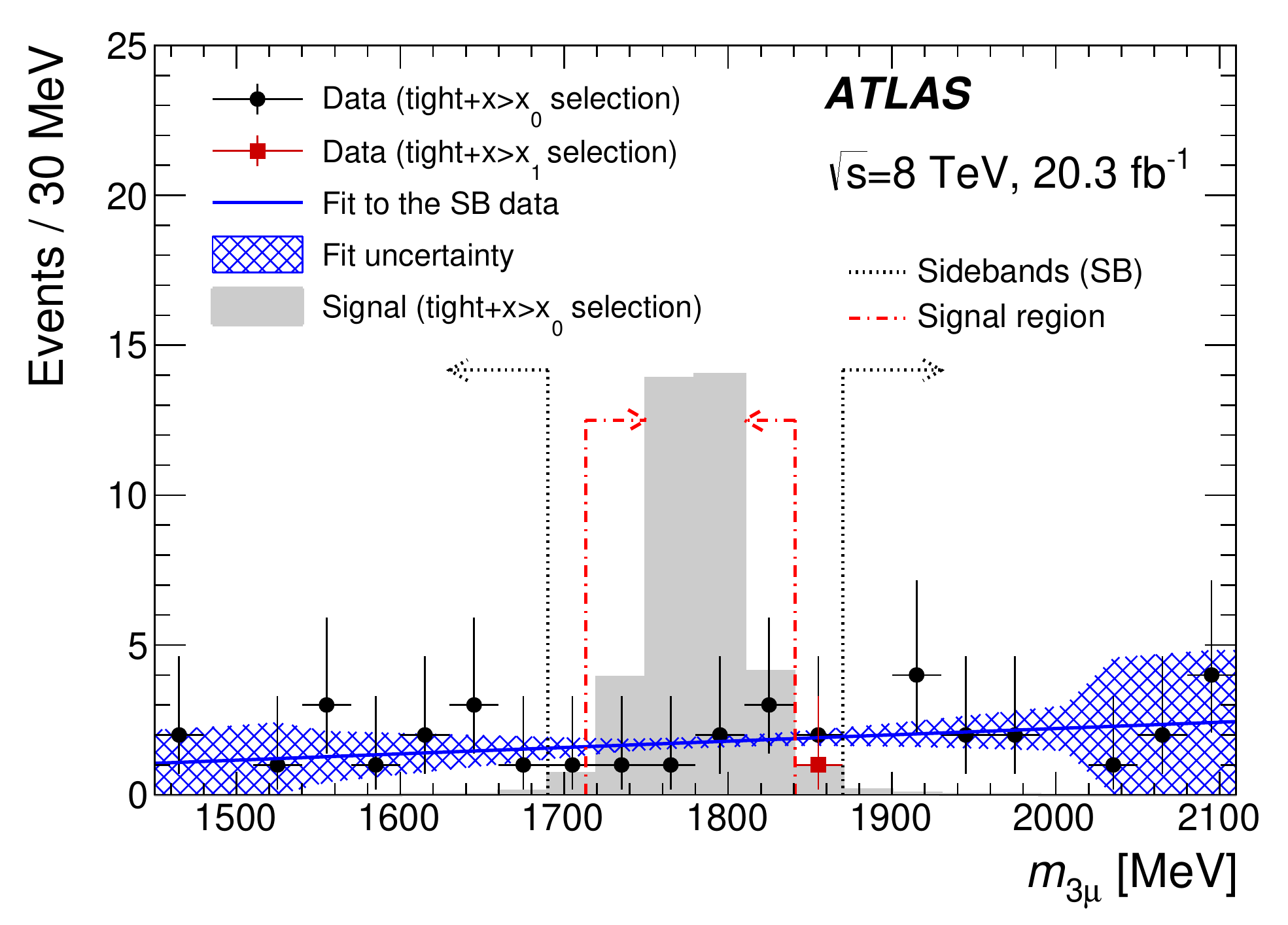}
\caption{The three-muon mass distribution in the range $[1450,2110]$~MeV shown for the \tightxzero selection by solid black circles and for the \tightxone selection by the solid red square. The sideband and signal regions are indicated by the arrows. The \tightxzero data are fit in the two sidebands simultaneously, excluding the events in the blinded region. The hatched area shows the uncertainty in the fit due to the SB range definition, the $x_0$ cut location and the fit function choice. The solid grey area shows the signal shape (obtained from MC simulation), normalised to the area of the data for the \tightxzero selection.}
\label{fig:SB}
\end{figure}

For any $x_1$ cut value above $x\sim 0.6$, where most of the signal is expected, the estimated \NbkgZero in the signal region can be then scaled down according to the ratio of the integrals of the BDT analytical function above and below this cut.
This ratio is denoted hereafter by $\mathcal{R}(x_1)$.
The extrapolation procedure can be written as $\NbkgOne = \mathcal{R}(x_1)\times\NbkgZero$ where \NbkgOne is estimated in the signal region for \xone.

\subsection{Uncertainties and optimisation}
\label{sec:uncert_optim}
The sources of systematic uncertainty associated with the extrapolation procedure in the background estimation are the BDT and sideband fit function choice, the definition of the sideband ranges and the definition of $x_0$.
To estimate this uncertainty, each of these definitions and choices is varied individually while calculating \NbkgZero and $\mathcal{R}(x_1)$.
For each fit function (BDT and sideband), different parameterisations are considered.
In addition, to construct the variation of the \tightxzero sample with which the two fits are performed, nine different sideband range variations and ten different $x_0$ variations are used.
The fits, and consequently also the extrapolation procedure, are found to be stable against these variations.
The dominant uncertainty component is the impact on $\mathcal{R}(x_1)$ of varying the sideband ranges definition.
The differences from the nominal values of $\mathcal{R}(x_1)$ and \NbkgZero are summed in quadrature and are translated to uncertainties in \NbkgOne.
The systematic uncertainty associated with the extrapolation procedure used to obtain \NbkgOne increases with $x_1$ from $\sim 45\%$ at $x_1=0.6$ to $\sim 80\%$ at $x_1\simeq 1$.
The statistical uncertainty of \NbkgOne is $\sim 19\%$, independent of $x_1$.

The systematic uncertainty in the signal acceptance times efficiency has contributions from reconstruction (13.1\%), trigger (11\%) and MC modelling (4\%) as discussed in the previous sections.
In addition, there is a small (2.1\%) contribution due to jet and \etmis calibration.
The number of $\tau$ leptons produced via the \Wtn channel and its uncertainty (3.9\%) are estimated as described in Section~\ref{sec:data_mc}.
These uncertainties are independent of $x_1$ in the range of interest.

The BDT cut is optimised by minimising the expected upper limit on the branching fraction given in Eq.~(\ref{eq:BR}), where \Nsig becomes the upper limit on the number of observed events above the expected background level in a narrow region around the $\tau$ lepton mass.
The procedure is performed by varying $x_1$ between 0.6 and 1.0 in steps of 0.001 while extracting \NbkgOne and its associated errors as explained above.
To obtain the upper limit on \Nsig for each $x_1$ cut, a single-bin counting experiment is performed using the {\sc HistFitter}~\cite{Baak:2014wma} statistical framework, supplied with \NbkgOne and its uncertainties.
For compatibility with previous searches, the limit on \Nsig and on \BRtmmm is given at 90\% confidence level (CL).
In each iteration, \acceff is calculated for the specific $x_1$ cut using a signal sample that is different from the one used for the BDT training.

During the iterative optimisation process, the extrapolation of the number of events in the sideband region to high $x_1$ cuts using the BDT shape is tested against a cut-and-count procedure.
The two procedures are found to agree very well within the uncertainties, and the extrapolation procedure gives a more conservative result throughout the examined $x_1$ range.
The resulting optimal cut is at $x_1=0.933$.

\section{Results}
\label{sec:results}
Figure~\ref{fig:SB} shows the three-muon mass distributions in the full mass range, including the blinded region, for the \tightxone selection in red squares.
Only one event with a three-muon mass of 1860~MeV survives the selection in the full mass range (sideband and blinded regions).
This event is found in the range between the signal region and the right sideband region and it does not affect the background estimation or the observation in the signal region.

The event counts entering the different regions at the different steps of the analysis for signal and data are given in table~\ref{table:eventcounts}.
\begin{table}[!ht]
\begin{center}
\begin{tabular}{l|r|rr}
\hline
\hline
Phase        & Data SB  & Data SR & Signal MC SR   \\
\,           & \,       & \,      & [out of $2\times 10^5$]  \\
\hline                                             
\loose       & 2248	    & 580     & 12672          \\
\loosexzero  & 736	    & 203     & 12557          \\
\tight       & 42	    & 9       & 5503           \\
\tightxzero  & 28	    & 7       & 5501           \\
\tightxone   & 0	    & 0       & 4616           \\
\hline 
\hline
\end{tabular}
\end{center}
\caption[Event counts]{The event count for the different steps of the analysis in the sideband and signal regions. The signal sample used to evaluate the \acceff has $2\times 10^5$ events.}
\label{table:eventcounts}
\end{table}

The signal acceptance times efficiency is calculated from the second signal MC sample after applying the full \tightxone selection.
This selection corresponds to $\acceff=0.0231 \pm 0.0005_{{\rm jet}/\etmis} \pm 0.0009_{\rm modelling} \pm 0.0025_{\rm trig} \pm 0.0030_{\rm reco}$.
With this selection, the expected background yield is $\NbkgOne=0.193 \pm 0.131_{\rm syst} \pm 0.037_{\rm stat}$.
The systematic uncertainty on \acceff is dominated by the uncertainties in the reconstruction and trigger efficiency measurements.
The systematic uncertainty on \NbkgOne is dominated by the uncertainty in the extrapolation of the background from the \tightxzero selection to the \tightxone selection. 

The systematic uncertainties in \Nbkg are taken into account when calculating the limit on the number of signal events, \Nsig, via one nuisance parameter.
The systematic uncertainties in the product $\left(\acceff\right)\cdot\NWtn$ are summed in quadrature and taken into account as the uncertainty in the signal via one nuisance parameter when calculating the limit.
The expected (median) limit on the branching fraction for $\Nobs=\NbkgOne$ is $3.94\times 10^{-7}$ at 90\% CL.
No events are observed in the signal region and the observed limit on the branching fraction is therefore $3.76\times 10^{-7}$ at 90\% CL.

\section{Conclusions and outlook}
\label{sec:discussion}
This article presents a search with the ATLAS detector for neutrinoless \tmmm decays using 20.3~\ifb of 2012 LHC $pp$ collision data, utilising $\tau$ leptons produced in \Wtn decays.
No events are observed in the signal region for the final selection while $0.193 \pm 0.131_{\rm syst} \pm 0.037_{\rm stat}$ background events are expected.
This results in an observed (expected) upper limit of $3.76\times 10^{-7}$ ($3.94\times 10^{-7}$) on \BRtmmm at 90\% CL.
Although this limit is not yet competitive with searches performed at $B$-factories~\cite{Babar:2010ez,BelleLimit:2007} and at LHCb~\cite{LHCb:2014}, it demonstrates the potential of LHC data collected by ATLAS as a probe of lepton flavour violation in $\tau$ lepton decays.
This analysis utilises single $\tau$ lepton production in an environment very different from $B$-factories, which rely on $\tau$ lepton pair production in $e^+e^-$ collisions.
The method and sample presented here were used to improve the ATLAS muon trigger and reconstruction of low-\pT, collimated muons relevant to the \tmmm search.
The analysis is limited by the number of \Wtn decays and by the systematic uncertainty, which depends on the size of the data sample.
With the much larger data sets anticipated at \runii of the LHC, the sensitivity of ATLAS to lepton-flavour-violating decays will be improved significantly. 
\section*{Acknowledgements}
\label{sec:acknowldge}

We thank CERN for the very successful operation of the LHC, as well as the
support staff from our institutions without whom ATLAS could not be
operated efficiently.

We acknowledge the support of ANPCyT, Argentina; YerPhI, Armenia; ARC, Australia; BMWFW and FWF, Austria; ANAS, Azerbaijan; SSTC, Belarus; CNPq and FAPESP, Brazil; NSERC, NRC and CFI, Canada; CERN; CONICYT, Chile; CAS, MOST and NSFC, China; COLCIENCIAS, Colombia; MSMT CR, MPO CR and VSC CR, Czech Republic; DNRF and DNSRC, Denmark; IN2P3-CNRS, CEA-DSM/IRFU, France; GNSF, Georgia; BMBF, HGF, and MPG, Germany; GSRT, Greece; RGC, Hong Kong SAR, China; ISF, I-CORE and Benoziyo Center, Israel; INFN, Italy; MEXT and JSPS, Japan; CNRST, Morocco; FOM and NWO, Netherlands; RCN, Norway; MNiSW and NCN, Poland; FCT, Portugal; MNE/IFA, Romania; MES of Russia and NRC KI, Russian Federation; JINR; MESTD, Serbia; MSSR, Slovakia; ARRS and MIZ\v{S}, Slovenia; DST/NRF, South Africa; MINECO, Spain; SRC and Wallenberg Foundation, Sweden; SERI, SNSF and Cantons of Bern and Geneva, Switzerland; MOST, Taiwan; TAEK, Turkey; STFC, United Kingdom; DOE and NSF, United States of America. In addition, individual groups and members have received support from BCKDF, the Canada Council, CANARIE, CRC, Compute Canada, FQRNT, and the Ontario Innovation Trust, Canada; EPLANET, ERC, FP7, Horizon 2020 and Marie Sk{\l}odowska-Curie Actions, European Union; Investissements d'Avenir Labex and Idex, ANR, R{\'e}gion Auvergne and Fondation Partager le Savoir, France; DFG and AvH Foundation, Germany; Herakleitos, Thales and Aristeia programmes co-financed by EU-ESF and the Greek NSRF; BSF, GIF and Minerva, Israel; BRF, Norway; the Royal Society and Leverhulme Trust, United Kingdom.

The crucial computing support from all WLCG partners is acknowledged
gratefully, in particular from CERN and the ATLAS Tier-1 facilities at
TRIUMF (Canada), NDGF (Denmark, Norway, Sweden), CC-IN2P3 (France),
KIT/GridKA (Germany), INFN-CNAF (Italy), NL-T1 (Netherlands), PIC (Spain),
ASGC (Taiwan), RAL (UK) and BNL (USA) and in the Tier-2 facilities
worldwide.

\clearpage
\appendix

\printbibliography

\newpage
\begin{flushleft}
{\Large The ATLAS Collaboration}

\bigskip

G.~Aad$^\textrm{\scriptsize 86}$,
B.~Abbott$^\textrm{\scriptsize 114}$,
J.~Abdallah$^\textrm{\scriptsize 152}$,
O.~Abdinov$^\textrm{\scriptsize 11}$,
R.~Aben$^\textrm{\scriptsize 108}$,
M.~Abolins$^\textrm{\scriptsize 91}$,
O.S.~AbouZeid$^\textrm{\scriptsize 159}$,
H.~Abramowicz$^\textrm{\scriptsize 154}$,
H.~Abreu$^\textrm{\scriptsize 153}$,
R.~Abreu$^\textrm{\scriptsize 117}$
% ...
\bigskip
\\
% ...
$^{11}$ Institute of Physics, Azerbaijan Academy of Sciences, Baku, Azerbaijan\\
% ...
$^{86}$ CPPM, Aix-Marseille Universit{\'e} and CNRS/IN2P3, Marseille, France\\
% ...
$^{91}$ Department of Physics and Astronomy, Michigan State University, East Lansing MI, United States of America\\
% ...
$^{108}$ Nikhef National Institute for Subatomic Physics and University of Amsterdam, Amsterdam, Netherlands\\
% ...
$^{114}$ Homer L. Dodge Department of Physics and Astronomy, University of Oklahoma, Norman OK, United States of America\\
% ...
$^{117}$ Center for High Energy Physics, University of Oregon, Eugene OR, United States of America\\
% ...
$^{152}$ Institute of Physics, Academia Sinica, Taipei, Taiwan\\
$^{153}$ Department of Physics, Technion: Israel Institute of Technology, Haifa, Israel\\
$^{154}$ Raymond and Beverly Sackler School of Physics and Astronomy, Tel Aviv University, Tel Aviv, Israel\\
% ...
$^{159}$ Department of Physics, University of Toronto, Toronto ON, Canada\\
% ...
\end{flushleft}


\end{document}